\documentclass[11pt]{article}

\usepackage[top=1.0in,bottom=1.0in,left=1.0in,right=1.0in]{geometry}
\usepackage{bigints}
\usepackage{amsmath,amssymb,amsthm}
\usepackage{tabularx}
\usepackage{graphicx}
\usepackage{float}
\usepackage{stmaryrd} 
\usepackage{relsize} 
\usepackage{algorithm}
\usepackage{algpseudocode}
\usepackage{color}
\usepackage[dvipsnames]{xcolor}
\usepackage{setspace}
\usepackage{enumitem}
\usepackage[normalem]{ulem} 
\usepackage[font=footnotesize]{caption}
\usepackage{hyperref}
\hypersetup{
    colorlinks=true,
    linkcolor=magenta,
    urlcolor=blue,
    citecolor=blue
}
\usepackage{tikz}
\usetikzlibrary{arrows.meta}

\usepackage{mathtools}
\usepackage{etoolbox}
\DeclareFontFamily{U}{matha}{\hyphenchar\font45}
\DeclareFontShape{U}{matha}{m}{n}{
      <5> <6> <7> <8> <9> <10> gen * matha
      <10.95> matha10 <12> <14.4> <17.28> <20.74> <24.88> matha12
      }{}
\DeclareSymbolFont{matha}{U}{matha}{m}{n}
\DeclareFontSubstitution{U}{matha}{m}{n}

\DeclareFontFamily{U}{mathx}{\hyphenchar\font45}
\DeclareFontShape{U}{mathx}{m}{n}{
      <5> <6> <7> <8> <9> <10>
      <10.95> <12> <14.4> <17.28> <20.74> <24.88>
      mathx10
      }{}
\DeclareSymbolFont{mathx}{U}{mathx}{m}{n}
\DeclareFontSubstitution{U}{mathx}{m}{n}

\DeclareMathDelimiter{\vvvert}{0}{matha}{"7E}{mathx}{"17}
\DeclarePairedDelimiterX{\normi}[1]
  {\vvvert}
  {\vvvert}
  {\ifblank{#1}{\:\cdot\:}{#1}}

\newcommand{\nx}{s}
\newcommand{\nt}{n_{t}}
\newcommand{\nxx}{n}

\newcommand{\Act}{\mathcal{S}}
\newcommand{\Ham}{h}

\newcommand{\Imega}{\mathcal{I}}

\newcommand{\tr}{^{T}}
\newcommand{\mtr}{^{-T}}
\newcommand{\lin}{\text{lin}}

\newcommand{\psip}{\dot{\psi}}

\newcommand{\uu}{\boldsymbol{u}}
\newcommand{\xx}{\boldsymbol{x}}
\newcommand{\vv}{\boldsymbol{v}}
\newcommand{\qq}{\boldsymbol{q}}
\newcommand{\pp}{\boldsymbol{p}}
\newcommand{\qqs}{\qq^{\ast}}
\newcommand{\qqp}{\dot{\boldsymbol{q}}}
\newcommand{\qqpp}{\ddot{\boldsymbol{q}}}
\newcommand{\ppp}{\dot{\boldsymbol{p}}}
\newcommand{\pps}{\boldsymbol{p}^{\ast}}
\newcommand{\ff}{\boldsymbol{f}}
\newcommand{\mmu}{\boldsymbol{\mu}}
\newcommand{\rr}{\boldsymbol{r}}
\newcommand{\zz}{\boldsymbol{z}}
\newcommand{\zzp}{\dot{\zz}}
\newcommand{\zzs}{\zz^{\ast}}
\newcommand{\yy}{\boldsymbol{y}}
\newcommand{\yyp}{\dot{\yy}}
\newcommand{\yys}{\yy^{\ast}}
\newcommand{\vvarphi}{\boldsymbol{\varphi}}
\newcommand{\vvarphis}{\vvarphi^{\ast}}
\newcommand{\ppsi}{\boldsymbol{\psi}}
\newcommand{\ppsis}{\ppsi^{\ast}}
\newcommand{\ppsip}{\dot{\ppsi}}
\newcommand{\Psip}{\dot{\Psi}}
\newcommand{\bb}{\boldsymbol{b}}

\newcommand\rev[1]{#1}

\begin{document}

\title{An Efficient PGD Solver for Structural Dynamics Applications}

\author{Cl\'ement Vella\textsuperscript{a,}\thanks{Corresponding author}, Pierre Gosselet\textsuperscript{a}, Serge Prudhomme\textsuperscript{b} \\[0.1in]
\footnotesize \textsuperscript{a}LaMcube -- Univ. Lille, CNRS, Centrale Lille, UMR 9013 -- F-59000 Lille, France \\
\footnotesize \textsuperscript{b}Department of Mathematics and Industrial Engineering, Polytechnique Montr\'eal,\\
\footnotesize C.P. 6079, succ.\ Centre-ville, Montr\'eal, Qu\'ebec H3C~3A7, Canada
}

\makeatletter
\def\blfootnote{\gdef\@thefnmark{}\@footnotetext}
\makeatother

\blfootnote{\vspace{-0.1in} \\
\textit{E-mail addresses:}~\url{clement.vella@univ-lille.fr} (C.~Vella) \\
\phantom{\textit{E-mail addresses:}}~\url{pierre.gosselet@univ-lille.fr} (P.~Gosselet) \\
\phantom{\textit{E-mail addresses:}}~\url{serge.prudhomme@polymtl.ca} (S.~Prudhomme)
}

\date{}

\maketitle

\noindent\rule{\textwidth}{0.5pt}
\vspace{-1cm}
\section*{Abstract}

We propose in this paper a Proper Generalized Decomposition (PGD) solver for reduced-order modeling of linear elastodynamic problems. It primarily focuses on enhancing the computational efficiency of a previously introduced PGD solver based on the Hamiltonian formalism. The novelty of this work lies in the implementation of a solver that is halfway between Modal Decomposition and the conventional PGD framework, so as to accelerate the fixed-point iteration algorithm. Additional procedures such that Aitken's delta-squared process and mode-orthogonalization are incorporated to ensure convergence and stability of the algorithm. Numerical results regarding the ROM accuracy, time complexity, and scalability are provided to demonstrate the performance of the new solver when applied to dynamic simulation of a three-dimensional structure.

\vspace{.1in}
\noindent\textit{Keywords:} Model Reduction, Proper Generalized Decomposition, Hamiltonian Formulation, Symplectic Structure, Ritz Pairs

\noindent\rule{\textwidth}{0.5pt}

\section{Introduction}
\label{sect:introduction}

Despite remarkable progress achieved in Computational Sciences and Engineering over the past decades, it is still necessary to develop innovative numerical methods to simplify models and make them easier to interpret for researchers and engineers in design offices. In linear structural dynamics, Modal Decomposition~\cite{rixen} with truncation undoubtedly remains the most popular technique among engineering analysis tools. It relies on computing the eigenvectors to describe the natural response of a given system. Unfortunately, not all eigenvectors are necessarily relevant to obtain the structural response under external loads, or, conversely, it may introduce a large number of these vectors to describe the mechanical behavior, which is not desirable in reduced-order modeling. Alternative approaches have been proposed for model reduction in structural dynamics, such as the Proper Orthogonal Decomposition (POD) or the Proper Generalized Decomposition (PGD) approaches. The POD method~\cite{kerschen, bamerPOD, lu} has been successfully applied to linear and nonlinear structures subjected to transient load and can be viewed as an \textit{a posteriori} approach, in the sense that it takes the state of the full-order model at different time-steps as input, the so-called snapshots, in order to extract the dominant spatial and temporal modes in the data. By contrast, the PGD method~\cite{chinesta2013,boucinha2013, boucinha_these} constructs a reduced basis on-the-fly, eliminating the need for prior knowledge of the solution to the problem. In that respect, the PGD method is used as an \textit{a priori} approach and can be assimilated as a solver: one simultaneously solves the problem and constructs a reduced approximation subspace.

The strength of the PGD strategy resides in the way it reduces high-dimension problems into subproblems of lower dimensions. The theoretical complexity of PGD solvers also decreases compared to conventional solvers. Indeed, following~\cite{chinesta2013}, one observes that if a solution is sought for in a space of a given dimension~$d$, the complexity of conventional solvers scales exponentially with~$d$ while that of PGD solvers scales linearly. However, the performance of the PGD approach using a space-time separation for transient structural dynamics has often been considered unsatisfactory~\cite{boucinha2013, boucinha_these}. One reason is that the fixed-point algorithm employed by Galerkin-based PGD solvers tends to exhibit poor convergence, if it converges at all~\cite{boucinha2013}. In fact, it is open to question whether the PGD framework using space-time separability is suitable for solving the wave equation, or more generally, second-order hyperbolic problems. The authors in~\cite{boucinha2013, boucinha2014} have proposed an alternative to the Galerkin-based PGD, namely the minimal residual PGD, that would consistently converge, a proof of which is given in~\cite{ammar2010TC}. 

On another note, the PGD framework has also been extended to perform basic operations, such as divisions, or more complex operations, such as solving linear systems of algebraic equations, leveraging the principle of variable separation. The authors in~\cite{diez} have thus created a versatile toolbox for PGD algebraic operators, which has been used in a non-intrusive manner to solve parametric eigen problems arising, for instance, in automotive applications~\cite{cavaliere2022}. Furthermore, notable advancements on the development of the PGD framework have been achieved using separated representations with respect to the space and frequency variables~\cite{malik2018-1, quaranta2019}. Moreover, it provides a means to take into account the parametric variability of a system due to, for example, material properties or geometric topology.  In this respect, the advantage of PGD solvers seems clear when geometric or material parameter separation is at stake, offering a considerable reduction of the computational complexity~\cite{chinesta2013, boucinha_these}. However, the space-frequency formulation does not necessarily provide direct insights into the transient behavior of the system. While it can determine its response at specific frequencies, it may fail to accurately capture time-dependent loads or dynamical events.

This was the motivation of the previous work~\cite{vella2022}, in which we introduced a new space-time Galerkin-based PGD solver based on the Hamiltonian formalism, which leads to an algorithm that was shown to be more stable than the Galerkin-based solver mentioned above. The novelty of the solver lied in the implementation of procedures that ensure linear independence of the modes and stability of the reduced-order model while progressively computing the new modes. However, the relevance of a reduced-order modeling technique stems from its ability to exceed the computational efficiency of a conventional Finite Element model, while incurring a relatively low error with respect to the FE solution of the full model. So far, if space-time PGD solvers have demonstrated a satisfying level of accuracy with a rather low number of modes, their computational efficiency is far from being competitive~\cite{boucinha_these}. In this paper, we develop a novel space-time PGD solver with a focus on computational efficiency. The integration of the PGD strategy within the Hamiltonian formalism is revisited and we comment on the preservation of the symplectic structure \rev{on the time parameter} by the reduced model. The Aitken transformation~\cite{aitken} has subsequently been introduced to accelerate the convergence of the fixed-point algorithm. We will show that it significantly reduces the number of required iterations for convergence. Additionally, a new orthogonal projection, more robust than the one formerly implemented, is performed on the spatial modes to enforce their linear independence and ensure the stability of the algorithm. Yet, the computational cost of such solvers mainly depends on the problem with respect to the spatial variable, which needs to be assembled and factorized at each fixed-point iteration. An original approach has been developed to avoid having to repeatedly factorize matrices. It consists in pre-processing the eigen-pair approximations of the operators, namely the Ritz pairs~\cite{sorensen}, that provide a subspace in which the problem in space remains diagonal throughout the fixed-point iterations. In the manner of Modal Decomposition, all computations are then carried out in the subspace spanned by the Ritz vectors~\cite{gosselet}, hence drastically decreasing the computational burden while capturing using only a small number of modes most of the information from the full model. Numerical examples dealing with the dynamical behavior of a 3D structure will be presented in order to demonstrate the efficiency of the proposed approach.

The paper is organized as follows: in Section~\ref{sect:modelproblem}, we describe the model problem and its spatial Finite Element approximation. In Section~\ref{sect:hamiltonian_form}, we present the Hamiltonian formalism and its symplectic structure. The PGD approaches are described in Section~\ref{sect:PGD} along with the Aitken acceleration and the orthogonal projectors applied to the fixed-point algorithm, as well as the projection of the PGD approximation onto the subspace spanned by the Ritz vectors. The numerical experiments are presented in Section~\ref{sect:numerical_results} to illustrate the performance of the proposed approach. We finally provide some concluding remarks in Section~\ref{sect:conclusions}.

\section{Model problem}
\label{sect:modelproblem}

\subsection{Strong formulation}

The model problem we shall consider is that of elastodynamics in three dimensions under the assumption of infinitesimal deformation. Let $\Omega$ be an open bounded subset of $\mathbb{R}^{3}$, with Lipschitz boundary $\partial \Omega$, and let $\Imega = (0, T)$ denote the time interval. The boundary $\partial\Omega$ is supposed to be decomposed into two portions, $\partial\Omega_D$ and $\partial\Omega_N$, such that $\partial\Omega = \overline{\partial\Omega_D \cup \partial\Omega_N}$. The displacement field $u : \bar{\Omega} \times \bar{\Imega} \rightarrow \mathbb{R}^{3}$ satisfies the following partial differential equation:
\begin{equation}
\label{eq:elastodyn}
\rho \frac{\partial^{2} u}{\partial t^{2}} - \nabla \cdot \sigma(u) = f, \qquad \forall (x, t) \in \Omega \times \Imega,
\end{equation}
where, in the case of infinitesimal deformation, the stress tensor~$\sigma(u)$ and strain tensor~$\varepsilon(u)$ are given by:
\begin{alignat}{2}
\label{eq:constitutivelaw}
& \sigma(u) = \mathbb{E} : \varepsilon(u), 
&&\qquad \forall (x, t) \in \Omega \times \Imega, \\
\label{eq:smalldeformation}
& \varepsilon(u) = \frac{1}{2} \Big( \nabla u + \big( \nabla u \big)\tr \Big), 
&&\qquad \forall (x, t) \in \Omega \times \Imega,
\end{alignat}
and is subjected to the initial conditions:
\begin{alignat}{2}
\label{eq:ICu}
&u(x, 0) = u_{0}(x), 
&&\qquad \forall x \in \Omega, \\
\label{eq:ICv}
&\frac{\partial u}{\partial t}(x, 0) = v_{0}(x), 
&&\qquad \forall x \in \Omega,
\end{alignat}
as well as to the boundary conditions:
\begin{alignat}{2}
\label{eq:DirichletBC}
&u(x, t) = 0, &&\qquad \forall (x, t) \in \partial\Omega_{D} \times \Imega, \\
\label{eq:NeumannBC}
&\sigma(u) \cdot n = g_{N}(x, t), &&\qquad \forall (x, t) \in \partial\Omega_{N} \times \Imega.
\end{alignat}
The functions $f : {\Omega} \times {\Imega} \rightarrow \mathbb{R}^{3}$, $u_{0} : {\Omega} \rightarrow \mathbb{R}^{3}$, $v_{0} : {\Omega} \rightarrow \mathbb{R}^{3}$, and $g_{N} : \partial\Omega_{N} \times {\Imega} \rightarrow \mathbb{R}^{3} $ are supposed to be sufficiently regular to yield a well-posed problem. The medium occupied by $\bar{\Omega}$ is assumed to be isotropic, with density~$\rho$ and Lam\'e coefficients~$\lambda$,~$\mu$ (the material parameters could possibly vary in space). The constitutive equation~\eqref{eq:constitutivelaw}, written above in terms of the tensor of elasticity~$\mathbb{E}$, thus reduces to:
\[
\sigma(u) = 2 \mu \varepsilon( u ) + \lambda \mathrm{tr} \left( \varepsilon(u) \right) I_{3},
\]
where $I_{3} \in \mathbb{R}^{3 \times 3}$ is the identity matrix.
In the following, we will denote the first and second time derivatives by $\dot{u} = \partial u /\partial t$ and $\ddot{u} = \partial^2 u/\partial t^2$.

\subsection{Semi-weak formulation}

We consider here the semi-weak formulation with respect to the spatial variable in order to construct the discrete problem in space using the Finite Element method. Multiplying~\eqref{eq:elastodyn} by an arbitrary smooth function $u^{\ast}=u^{\ast}(x)$ and integrating over the whole domain $\Omega$, one obtains:
\begin{equation}
\label{eq:baseweak_form}
\int_{\Omega}{ \rho \ddot{u} \cdot u^{\ast} - \left( \nabla \cdot \sigma(u) \right) \cdot u^{\ast} ~dx } 
= \int_{\Omega}{ f \cdot u^{\ast} ~dx }, \qquad \forall t \in \mathcal I.
\end{equation}
By virtue of $-\left( \nabla \cdot \sigma(u) \right) \cdot u^{\ast} 
= \sigma(u) : \nabla u^{\ast} 
- \nabla \cdot \left( \sigma( u ) \cdot u^{\ast} \right)$, 
Eq.~\eqref{eq:baseweak_form} can be recast as:
\[
\int_{\Omega}{ \rho \ddot{u} \cdot u^{\ast} + \sigma(u) : \nabla u^{\ast} ~dx } 
= \int_{\Omega}{  \nabla \cdot \left(\sigma(u)  \cdot u^{\ast} \right) ~dx} 
+ \int_{\Omega}{ f \cdot u^{\ast} ~dx }, \qquad \forall t \in \mathcal I.
\]
Since $\sigma(u)$ is a symmetric tensor:
\[
\sigma(u) : \nabla u^{\ast} = \sigma(u) : \varepsilon(u^{\ast}),
\]
and substituting the constitutive equation for $\sigma(u)$, one gets:
\[
\sigma(u) : \varepsilon(u^{\ast}) = \left( \mathbb{E} : \varepsilon(u) \right) : \varepsilon(u^{\ast}) =  \varepsilon(u) : \mathbb{E} : \varepsilon(u^{\ast}).
\]
Applying the divergence theorem and the boundary conditions, and choosing the test function such that $u^\ast = 0$ on $\partial\Omega_D$, 
the semi-discrete formulation of the problem then reads: Find $u=u(\cdot,t) \in V$, for all $t\in \bar{\mathcal I}$, such that:
\begin{equation}
\label{eq:weak_form}
\int_{\Omega}{ \rho \ddot{u} \cdot u^{\ast} + \varepsilon(u) : \mathbb{E} : \varepsilon(u^{\ast}) ~dx } 
= \int_{\Omega}{ f \cdot u^{\ast} ~dx }
+ \int_{\partial \Omega_{N}}{ g_{N} \cdot u^{\ast} ~dx }, 
\qquad \forall u^{\ast} \in V, \quad \forall t \in \mathcal I,
\end{equation}
and:
\begin{alignat}{2}
&u(x, 0) = u_{0}(x), 
&&\qquad \forall x \in \Omega, \\
&\frac{\partial u}{\partial t}(x, 0) = v_{0}(x), 
&&\qquad \forall x \in \Omega,
\end{alignat}
where $V$ is the vector space of vector-valued functions defined on $\Omega$:
\[
V = \left\{ v \in \left[ H^{1}(\Omega) \right]^{3}:\ v = 0~\text{on}~\partial \Omega_{D} \right\}.
\]

\subsection{Spatial discretization}
\label{subsect:fem_ref}

We partition the domain into $N_e$ elements $K_e$ such that $\overline{\Omega} = \cup_{e=1}^{N_e} K_e$ and $\text{Int}(K_i) \cap \text{Int}(K_j) = \varnothing$, $\forall i,j=1,\ldots,N_e$, $i\neq j$. We then associate with the mesh the finite-dimensional Finite Element space~$W^h$, $\text{dim}\ W^h = \nx$, of scalar-valued continuous and piecewise polynomial functions defined on $\Omega$, that is:
\begin{equation*}
W^h = \left\{ v_{h}: \Omega \rightarrow \mathbb R:\ v_{h}|_{K_{e}} \in \mathbb P_{k}(K_{e}),\ e = 1, \ldots, N_{e} \right\}, 
\end{equation*}
where $\mathbb P_k(K_e)$ denotes the space of polynomial functions of degree $k$ on $K_e$. Let $\phi_i$, $i=1,\ldots,\nx$, denote the basis functions of $W^h$, i.e.\ $W^{h} = \text{span}\{\phi_{i}\}$. We then introduce the finite element subspace $V^h$ of $V$ such that:
\[
V^h = \big[ W^h \big]^3 \cap V,
\]
and search for finite element solutions $u_h=u_h(\cdot, t) \in V^h$, $\forall t \in \bar{\mathcal I}$, in the form:
\begin{equation*}
u_{h}(x, t) = \sum_{j = 1}^{\nx} q_{j}(t) \phi_{j}(x),
\end{equation*}
where the vectors of degrees of freedom, $q_{j} \in \mathbb{R}^{3}$, depend on time. We introduce the set of $\nxx=3\nx$ vector-valued basis functions as:
\[
\chi_{3i-2}(x) = \begin{bmatrix} \phi_i(x) \\ 0 \\ 0 \end{bmatrix}, \quad
\chi_{3i-1}(x) = \begin{bmatrix} 0 \\ \phi_i(x) \\ 0 \end{bmatrix}, \quad
\chi_{3i}(x) = \begin{bmatrix} 0 \\ 0 \\ \phi_i(x) \end{bmatrix}, \qquad i =1,\ldots,\nx.
\]
Using the Galerkin method, the Finite Element problem thus reads:
\begin{align*}
&\text{Find $u_{h}(\cdot,t) \in V^{h}$, such that} \\
&\qquad\quad
\int_{\Omega}{ \rho \chi_{i}(x) \cdot \ddot{u}_{h}(x,t) + \varepsilon(\chi_i)(x) : \mathbb{E} : \varepsilon(u_h) (x,t) \, dx} \\
&\hspace{1.0in} 
= \int_{\Omega}{ \chi_i(x)\cdot f(x,t) \, dx} 
+ \int_{\partial\Omega_N}{ g_N (x,t)\cdot \chi_i(x) ~ds} , 
\quad \forall i=1,\ldots,\nxx, \quad \forall t \in \Imega, \\
&\text{satisfying the initial conditions} \nonumber \\
&\hspace{1.0in} 
u_{h}(x,0) = u_{0,h}(x), \quad \forall x \in \Omega, \\
&\hspace{1.0in} 
\dot{u}_h(x,0) = v_{0,h}(x), \quad \forall x \in \Omega,
\end{align*}
where $u_{0,h}$ and $v_{0,h}$ are interpolants or projections of $u_0$ and $v_0$ in the space $V^h$. The above problem can be conveniently recast in compact form as:
\begin{align}
\label{eq:discr_elasto}
M \qqpp(t) + K \qq(t) &= \ff(t), \qquad \forall t \in \Imega, \\
\label{eq:discr_ICu}
\qq(0) &= \uu_{0}, \\
\label{eq:discr_ICv}
\qqp(0) &= \vv_{0},
\end{align}
where $M$ and $K$ are the global mass and stiffness matrices, respectively, both being symmetric and positive definite:
\begin{equation*}
M_{ij} = \int_{\Omega}{ \rho \chi_{i}\cdot \chi_{j} ~dx}, \qquad
K_{ij} = \int_{\Omega}{ \varepsilon(\chi_i) : \mathbb{E} : \varepsilon(\chi_j) ~dx}, \qquad
\forall i,j=1,\ldots,\nxx,
\end{equation*}
$\ff(t)$ is the load vector at time $t$ whose components are given by:
\begin{equation*}
f_{i}(t) = \int_{\Omega}{ \chi_i(x) \cdot f(x,t) \, dx} + \int_{\partial\Omega_N}{ \chi_i(x)\cdot g_N (x,t) ~ds }, \qquad \forall i=1,\ldots,\nxx,
\end{equation*}
$\qq(t)$ is the global vector of degrees of freedom:
\begin{equation*}
\qq(t) = \begin{bmatrix} q_{1}(t) & \ldots & q_{\nx}(t) \end{bmatrix}\tr,
\end{equation*}
and $\uu_{0}$ and $\vv_{0}$ are the initial vectors:
\begin{equation*}
\begin{aligned}
& \uu_{0} = \begin{bmatrix} u_{0,1} & \ldots & u_{0,\nx} \end{bmatrix}\tr, \\
& \vv_{0} = \begin{bmatrix} v_{0,1} & \ldots & v_{0,\nx} \end{bmatrix}\tr. 
\end{aligned}
\end{equation*}
Note that $u_{0,i} \in \mathbb{R}^{3}$ and $v_{0,i} \in \mathbb{R}^{3}$, $i=1,\ldots,\nx$, are vectors whose components are the initial displacements and velocities in the three spatial directions. 

\section{The Hamiltonian formalism}
\label{sect:hamiltonian_form}

\subsection{Hamilton's Weak Principle}

The Hamiltonian formalism consists in modeling the motion of the system along a trajectory in the phase space by introducing the generalized coordinates $\qq$ and their generalized (or conjugate) momenta $\pp$ as independent variables. For the problem at hand, the Hamiltonian functional~$\Ham$ reads:
\begin{equation}
\label{eq:hamiltonian}
\Ham(\qq, \pp, t) = \frac{1}{2} \qq\tr K \qq + \frac{1}{2} \pp\tr M^{-1} \pp - \qq\tr \ff.
\end{equation}
Given the Hamiltonian functional~$\Ham$ of the system, the action functional, denoted by $\Act[\qq, \pp]$, is defined as: 
\[
\mathcal{S}[\qq, \pp] = \int_{\Imega}{\qqp\tr \pp - \Ham(\qq, \pp, t) ~dt}
\]
The Hamilton's Weak Principle then states that the trajectory $(\qq, \pp)$ of the system in the phase space should satisfy:
\[
\Act'[\qq, \pp](\qqs, \pps) = \left[ {\qqs}\tr \pp \right]_{0}^{T},
\]
where $\Act'[\qq, \pp](\qqs, \pps)$ denotes the G\^ateaux derivative of $\Act[\qq, \pp]$ with respect to a variation $(\qqs, \pps) \in \mathcal{Z} \times \mathcal{Z}$ such that:
\[
\mathcal Z = \left\{ \vv \in [C^1(\bar\Imega)]^{n};\ \vv(0) = 0 \right\}.
\]
After Gâteaux derivation and integration by parts with respect to time, we get:
\[ 
\int_{\Imega}{ {\pps}\tr \qqp - {\qqs}\tr \ppp - {\qqs}\tr K \qq - {\pps}\tr M^{-1} \pp + {\qqs}\tr \ff ~dt } = 0, \qquad \forall (\qqs, \pps) \in \mathcal Z \times \mathcal Z,
\]
that is, 
\[
\int_{\Imega}{ {\qqs}\tr \ppp - {\pps}\tr \qqp + {\qqs}\tr K \qq + {\pps}\tr M^{-1} \pp~ dt} = \int_{\Imega}{ {\qqs}\tr \ff ~dt}, \qquad \forall (\qqs, \pps) \in \mathcal Z \times \mathcal Z,
\]
or, equivalently,
\begin{equation}
\label{eq:ham_weak}
\begin{aligned}
&\int_{\Imega}{ {\qqs}\tr \left( \ppp + K \qq \right) ~dt} = \int_{\Imega}{ {\qqs}\tr \ff ~dt}, &&\qquad \forall \qqs \in \mathcal Z,\\
&\int_{\Imega}{ {\pps}\tr \left( \qqp - M^{-1} \pp \right) ~dt} = 0,
&&\qquad \forall \pps \in \mathcal Z.\\
\end{aligned}  
\end{equation}
The last weak formulation of~\eqref{eq:ham_weak} leads the so-called Hamilton's equations:
\begin{equation}
\label{eq:hamilton}
\begin{aligned}
\ppp + K \qq &= \ff, \\
\qqp - M^{-1} \pp &= \boldsymbol{0}.
\end{aligned} 
\end{equation}
This formulation is consistent with~\eqref{eq:discr_elasto} in the sense that if we differentiate with respect to time the second equation and substitute $\ppp$ for the expression in the first equation, we do exactly recover~\eqref{eq:discr_elasto}.

\subsection{Symplectic structure}

Let us introduce $\zz \in \mathcal Z^2$ that vertically concatenates $\qq$ and $\pp$ such that:
\[
\zz = \begin{bmatrix}
\qq \\ \pp
\end{bmatrix}.
\]
The gradient of the Hamiltonian~\eqref{eq:hamiltonian} then reads:
\[
\nabla_{\!z}\, \Ham 
= \begin{bmatrix} \nabla_{\!q}\, \Ham \\ \nabla_{\!p}\, \Ham \end{bmatrix} 
= \begin{bmatrix} K \qq - \ff \\ M^{-1} \pp \end{bmatrix}.
\]
In the symplectic framework, the dynamics of the structure is modeled by the trajectory in the symplectic vector space $(\mathbb{R}^{2\nxx}, \omega)$ of dimension $2\nxx$ for linear systems, where $\omega$ is the so-called symplectic form defined as:
\[
\forall \zz = \begin{bmatrix} \qq \\ \pp \end{bmatrix} \in \mathbb{R}^{2\nxx}, \quad 
\forall \zz' = \begin{bmatrix} \qq' \\ \pp' \end{bmatrix} \in \mathbb{R}^{2\nxx}, \quad 
\omega(\zz, \zz') = \qq\tr \pp' - {\qq'}\tr \pp = \zz\tr J_{2\nxx} \zz',
\]
with $J_{2\nxx}$ the skew-symmetric operator such that:
\[
J_{2\nxx} = \begin{bmatrix}
\phantom{-} 0 & I_{\nxx} \\ - I_{\nxx} & 0
\end{bmatrix},
\]
and $J_{2\nxx}^{2} = - I_{2\nxx}$. It is then possible to recast~\eqref{eq:hamilton} as:
\[
\zzp = \nabla^{\omega} h,
\]
where $\nabla^{\omega} = J_{2\nxx} \nabla_{\!z}$ is defined as the symplectic gradient. The Hamiltonian can be written as a sum of a quadratic form on $\mathbb{R}^{2\nxx}$ and the external energy term:
\[
h(\zz, t) = \frac{1}{2} \zz\tr H \zz - \zz\tr \ff_{z},
\]
with $H$ the Hessian operator of $\Ham$ and $\ff_{z}$ such that:
\[
H = \begin{bmatrix}
K & 0 \\ 0 & M^{-1}
\end{bmatrix},  \qquad \ff_{z} = \begin{bmatrix} \ff \\ 0 \end{bmatrix}.
\]
It follows that one can rewrite the weak formulation~\eqref{eq:ham_weak} as:
\begin{equation}
\label{eq:condham_weak}
\int_{\Imega}{ {\zzs}\tr \left[ J_{2\nxx} \zzp + H \zz \right]~ dt} 
= \int_{\Imega}{ {\zzs}\tr \ff_{z} ~dt}, \qquad \forall \zzs \in \mathcal Z^{2}. 
\end{equation}

We now introduce the notion of symplectic mapping. A symplectic mapping is a linear transformation that preserves the symplectic form $\omega$, i.e.:
\[
A \in \mathbb{R}^{2\nxx \times 2\nxx}\ \text{is symplectic if} \quad 
\omega(A \zz, A \zz') = \omega(\zz, \zz'),\quad 
\forall (\zz, \zz') \in \mathbb{R}^{2\nxx} \times \mathbb{R}^{2\nxx}.
\]
As a consequence, such a mapping $A$ verifies:
\[
A\tr J_{2\nxx} A = J_{2\nxx}.
\]
The notion can actually be generalized to rectangular matrices with the symplectic Stiefel manifold, denoted $S_{p}(2r, 2\nxx)$, such that:
\begin{equation}
\label{eq:def_stiefel}
S_{p}(2r, 2\nxx) = \left\{ A \in \mathbb{R}^{2\nxx \times 2r}:\  A\tr J_{2\nxx} A = J_{2r} \right\} \end{equation}
Let $(\mathbb{R}^{2r}, \gamma)$ be a symplectic vector space, $A \in S_{p}(2r, 2\nxx)$ a symplectic mapping, and $\yy \in \mathbb{R}^{2r}$ such that $\xx = A \yy$. One can define a Hamiltonian for $\yy$:
\[
g(\yy) = \frac{1}{2} \yy\tr G \yy - \yy\tr \ff_{y},
\]
with $G$ its Hessian operator and $\ff_{y}$ the projection of the external loads on the symplectic subspace (in the case $r\leqslant n$), such that:
\[
G = A\tr H A, \quad \text{and} \quad \ff_{y} = A\tr \ff_{z}.
\]
The preservation of the symplectic structure implies that $\yy$ is governed by Hamilton's canonical equations, expressed hereinafter in terms of $\gamma$ (symplectic form on $\mathbb{R}^{2r}$) and $g$ such that:
\[
\yyp = \nabla^{\gamma} g,
\]
with $\nabla^{\gamma} = J_{2r} \nabla_{\! y}$ and Hamilton's weak principle~\eqref{eq:ham_weak} then reads:
\[
\int_{\Imega}{ {\yys}\tr \left[ J_{2r} \yyp + G \yy \right]~ dt} 
= \int_{\Imega}{ {\yys}\tr \ff_{y} ~dt}, \qquad \forall \yys \in \mathbb{R}^{2r}.
\]

\subsection{Discretization in time of the Hamiltonian problem}
\label{subsect:time_disc}

The time domain $\Imega$ is divided into $\nt$ subintervals $\Imega^{i} = \left[ t^{i-1},t^{i} \right]$, $i=1,\ldots,\nt$, of size $h_{t}=t^{i} - t^{i-1}$. The Crank-Nicolson method is then applied to~\eqref{eq:hamilton} as detailed in the previous work~\cite{vella2022}. The solutions given by the FEM in space, integrated with Crank-Nicolson in time, will be used as reference solutions when assessing the results of the PGD solvers.

Although not the primary focus of this article, we acknowledge the relevance of symplectic integrators in the case of Hamiltonian mechanics. These integrators are particularly robust to compute long-time evolution of Hamiltonian systems~\cite{lew, razafindralandy, simo}. In addition, the preservation of the symplectic structure by the reduced model is the subject of numerous studies~\cite{babakmaboudi, buchfink, peng}. We will also discuss this property \rev{on the time parameter} with respect to our PGD solver in Section~\ref{subsect:tempuptsympl}.

\section{PGD reduced-order modeling}
\label{sect:PGD}

The proper-generalized decomposition method applied within the Hamiltonian framework aims at approximating both the generalized coordinates $\qq$ and their generalized momenta $\pp$ in separated form. We are thus searching for a space-time separated representation of $\zz$ as:
\[
\zz(t) \simeq \zz_{m}(t) 
= \sum_{i = 1}^{m}{ \Phi_{i} \ppsi_{i}(t) }
= \sum_{i = 1}^{m}{ \Psi_{i}(t) \vvarphi_{i} },
\]
with:
\[
\begin{aligned}
& \Phi_{i} = \begin{bmatrix} \vvarphi_{i}^{q} & 0 \\ 0 & \vvarphi_{i}^{p} \end{bmatrix}, 
\quad 
&\ppsi_{i} = \begin{bmatrix} \psi_{i}^{q} \\ \psi_{i}^{p} \end{bmatrix}, \\
& \Psi_{i} = \begin{bmatrix} \psi_{i}^{q} I_{\nxx} & 0 \\ 0 & \psi_{i}^{p} I_{\nxx} \end{bmatrix}, 
\quad 
&\vvarphi_{i} = \begin{bmatrix} \vvarphi_{i}^{q} \\ \vvarphi_{i}^{p} \end{bmatrix},
\end{aligned}
\]
where $\Phi_{i}$ is a $(2 n \times 2)$ matrix and $\ppsi_{i}$ a $(2 \times 1)$ vector while $\Psi_{i}$ is a $(2 n \times 2 n)$ matrix and $\vvarphi_{i}$ a $(2 n \times 1)$ vector. The two notations are mathematically equivalent and convenient whether the weak formulation is solved for $\vvarphi$ (spatial problem) or $\ppsi$ (temporal problem). The vector-valued functions $(\vvarphi_{i}^{q})_{1 \leqslant i \leqslant m}$ and $(\vvarphi_{i}^{p})_{1 \leqslant i \leqslant m}$ provide the spatial bases for the generalized coordinates and conjugate momenta, respectively:
\[
\begin{aligned}
&\qq(t) \approx \qq_{m}(t) = \sum_{i = 1}^{m}{ \vvarphi_{i}^{q} \psi_{i}^{q}(t) }, \\
&\pp(t) \approx \pp_{m}(t) = \sum_{i = 1}^{m}{ \vvarphi_{i}^{p} \psi_{i}^{p}(t) }.
\end{aligned}
\]
For the sake of clarity in the presentation, we shall drop from now on the subscript $i$ and write the decomposition of rank~$m$ of $\zz$ as:
\[
\zz_{m}(t) = \zz_{m - 1}(t) + \Phi \ppsi(t), 
\qquad \text{or} \qquad 
\zz_{m}(t) = \zz_{m - 1}(t) + \Psi(t) \vvarphi.
\]
The approach considered here is the so-called greedy rank-one update algorithm, where the separated representation is computed progressively by adding one pair of modes $\vvarphi$ and $\ppsi$ at each enrichment. The goal in this section is to construct the separated spatial and temporal problems that satisfy the enrichment modes $\vvarphi$ and $\ppsi$, the new unknowns of the problem, assuming that the previous iterate $\zz_{m - 1}$ has already been calculated.

\subsection{Fixed-point strategy}

Computing a separated representation of $\qq$ and $\pp$ demands an adequate solution strategy of the weak formulation~\eqref{eq:condham_weak}. Substituting the trial solution $\zz_{m}$ for $\zz$ in~\eqref{eq:condham_weak} leads to a non-linear formulation for the modes $\vvarphi$ and $\ppsi$. Several iterative schemes could be used to solve such a problem. The fixed point algorithm is considered here, which proceeds as follows:
\begin{enumerate}
\item Solve~\eqref{eq:condham_weak} for $\vvarphi$ with $\ppsi$ known. This step is referred to as the spatial problem and is written in a generic format as:
\begin{equation}
\label{eq:genspace}
A(\ppsi) \vvarphi = \bb(\ppsi, \zz_{m - 1}),
\end{equation}
where the matrix $A(\ppsi)$ and vector $\bb(\ppsi, \zz_{m - 1})$ will be specified in Section~\ref{subsubsect:space_pgd}.
More precisely, in order to enhance robustness, we propose to force the new spatial mode to preserve the linear independence of the spatial bases $(\vvarphi_{i}^{q})_{1 \leqslant i \leqslant m}$ and $(\vvarphi_{i}^{p})_{1 \leqslant i \leqslant m}$, which can formally be written as:
\[
 \vvarphi = P_m A(\ppsi)^{-1} \bb(\ppsi, \zz_{m - 1})
\]
where $P_m$ is a projector that is orthogonal to the subspace spanned by previous mode (for a well chosen inner product).
\item Solve~\eqref{eq:condham_weak} for $\ppsi$ with $\vvarphi$ known. The temporal problem corresponds to the system of first-order differential equations:
\begin{equation}
\label{eq:gentime}
\ppsip = f_{_\mathcal{T}}(\ppsi, \vvarphi, \zz_{m - 1}),
\end{equation}
where the vector-valued function $f_{_\mathcal{T}}$ will be explicitly provided in Section~\ref{subsubsect:time_pgd}.
\end{enumerate}
Steps~1 and~2 are repeated until a convergence criterion is fulfilled. It is noteworthy that~\eqref{eq:genspace} is a linear system of size $2 \nxx$ associated with the space discretization, similar to that of a steady-state FEM problem. Eq.~\eqref{eq:gentime} is a system of two first order scalar ordinary differential equations in time, solved for $\psi_{q}$ and $\psi_{p}$. Both problems are described in the next sections.

\subsubsection{Problem in space}
\label{subsubsect:space_pgd}

We assume that $\ppsi$ is known and search for the new spatial mode $\vvarphi$. We substitute $\zz_{m - 1} + \Psi \vvarphi$ for $\zz$ in~\eqref{eq:condham_weak} and choose test functions in the form $\zzs = \Psi \vvarphis$. Equation~\eqref{eq:condham_weak} reduces to:
\[
\int_{\Imega}{ {\vvarphis}\tr \Psi\tr \left( J_{2 n} \Psip \vvarphi + H \Psi \vvarphi \right)~ dt } 
= \int_{\Imega}{ {\vvarphis}\tr \Psi\tr \left( \ff - J_{2 n} \zzp_{m - 1} - H \zz_{m - 1} \right)~ dt }, 
\qquad \forall \vvarphis \in \mathbb R^{2n},
\]
which, since $\vvarphis$ and $\vvarphi$ are independent of time, can be rewritten as:
\[
{\vvarphis}\tr \left[ \int_{\Imega}{ \Psi\tr J_{2 n} \Psip + \Psi\tr H \Psi~ dt } \right] \vvarphi 
= {\vvarphis}\tr \left[ \int_{\Imega}{ \Psi\tr \left( \ff - J_{2 n} \zzp_{m - 1} - H \zz_{m - 1} \right)~ dt } \right], 
\qquad \forall \vvarphis \in \mathbb R^{2n}.
\]
This leads to the following linear system:
\begin{equation}
A_{\mathcal{S}} \vvarphi = \bb_{\mathcal{S}}, \label{eq:space_pgd}
\end{equation}
with:
\[
\begin{aligned}
A_{\mathcal{S}} &= \left[ \int_{\Imega}{ \Psi\tr J_{2 \nxx} \Psip + \Psi\tr H \Psi~ dt } \right] = \begin{bmatrix}
k_{t} K & c_{t} I_{\nxx} \\
d_{t} I_{\nxx} & m_{t} M^{-1}
\end{bmatrix}, \\[.1in]
\bb_{\mathcal{S}} &= \int_{\Imega}{ \Psi\tr \left( \ff_{z} - J_{2 n} \zzp_{m - 1} - H \zz_{m - 1} \right)~ dt },
\end{aligned}
\]
and \rev{(see Appendix \ref{appendix:timeop} for the explicit form of the time operators)}:
\[
\begin{aligned}
k_{t} &= \left(\int_{\Imega}{ \psi_{q}^{2}~ dt } \right), \\
c_{t} &= \left(\int_{\Imega}{ \psi_{q} \psip_{p}~ dt } \right), \\
d_{t} &= - \left(\int_{\Imega}{ \psip_{q} \psi_{p}~ dt } \right) = c_{t} - \psi_{q}(T) \psi_{p}(T), \\
m_{t} &= \left(\int_{\Imega}{ \psi_{p}^{2}~ dt } \right).
\end{aligned}
\]
The operator $M^{-1}$ is not computed explicitly. Instead, the Schur complement of $M^{-1}$ in $A_{\mathcal{S}}$ is considered. Equation~\eqref{eq:space_pgd} can thus be expanded as:
\[
\begin{aligned}
k_{t} K\ \vvarphi_{q} & \hspace{.1in} + &        c_{t}\ & \vvarphi_{p} = \bb_{q}, \\
  d_{t}\ \vvarphi_{q} & \hspace{.1in} + & m_{t} M^{-1}\ & \vvarphi_{p} = \bb_{p},
\end{aligned}
\]
so that:
\begin{align}
\label{eq:space_pgd1}
\left[ m_{t} k_{t} K - c_{t} d_{t} M \right] \vvarphi_{q} &= m_{t} \bb_{q} - c_{t} M \bb_{p}, \\
\label{eq:space_pgd2}
\vvarphi_{p} &= \frac{1}{m_{t}} M \left[ \bb_{p} - d_{t} \vvarphi_{q} \right].
\end{align}
Therefore, the solution of~\eqref{eq:space_pgd} amounts to solving~\eqref{eq:space_pgd1} for $\vvarphi_{q}$ by factorization of the sparse symmetric matrix:
\begin{equation}
\label{eq:matrixAq}
A_{q} = m_{t} k_{t} K - c_{t} d_{t} M,
\end{equation}
and inserting the solution $\vvarphi_{q}$ into~\eqref{eq:space_pgd2} to determine~$\vvarphi_{p}$.

For a given $m$\textsuperscript{th} enrichment, the spatial modes $\vvarphi_{q}$ and $\vvarphi_{p}$ are subsequently projected to ensure that any new mode is searched in a direction that is orthogonal to the subspaces generated by the previous modes, respectively $(\vvarphi_{i}^{q})_{1 \leqslant i \leqslant m - 1}$ and $(\vvarphi_{i}^{p})_{1 \leqslant i \leqslant m - 1}$. At any given $m$, we want $(\vvarphi_{i}^{q})_{1 \leqslant i \leqslant m}$ and $(\vvarphi_{i}^{p})_{1 \leqslant i \leqslant m}$ to be orthogonal with respect to $K$ and $M^{-1}$, respectively. Let $S_{q}$ and $S_{p}$ be defined as:
\[
\begin{aligned}
S_{q} &= \begin{bmatrix}
\vvarphi_{1}^{q} & \ldots & \vvarphi_{m - 1}^{q}
\end{bmatrix}, \\
S_{p} &= \begin{bmatrix}
\vvarphi_{1}^{p} & \ldots & \vvarphi_{m - 1}^{p}
\end{bmatrix}.
\end{aligned}
\]
A classical approach consists in using the orthogonal projections:
\[
\begin{aligned}
P_{q} &= I_{\nxx} - S_{q} \left( S_{q}\tr K S_{q} \right)^{-1} S_{q}\tr K, \\
P_{p} &= I_{\nxx} - S_{p} \left( S_{p}\tr M^{-1} S_{p} \right)^{-1} S_{p}\tr M^{-1}.
\end{aligned}
\]
At any enrichment step, the previous modes $(\vvarphi_{i}^{q})_{1 \leqslant i \leqslant m - 1}$ and $(\vvarphi_{i}^{p})_{1 \leqslant i \leqslant m - 1}$ are orthogonal and normalized with respect to $K$ and $M^{-1}$, respectively. Thus, the projectors above simplify as:
\[
\begin{aligned}
P_{q} &= I_{\nxx} - S_{q} S_{q}\tr K, \\
P_{p} &= I_{\nxx} - S_{p} S_{p}\tr M^{-1}.
\end{aligned}
\]
Therefore, if we denote by $\vvarphi_{q}^{\circ}$ and $\vvarphi_{p}^{\circ}$ the modes initially obtained from Eqs.~\eqref{eq:space_pgd1} and~\eqref{eq:space_pgd2} and by $\vvarphi_{q}$ and $\vvarphi_{p}$ the modes that one retains after orthonormalization, the procedure reads:
\[
\begin{aligned}
&\vvarphi_{q}^{\perp} = P_{q} \vvarphi_{q}^{\circ}, \qquad 
&& \vvarphi_{p}^{\perp} = P_{p} \vvarphi_{p}^{\circ}, \\
&\vvarphi_{q} = \frac{\vvarphi_{q}^{\perp}}{\sqrt{{\vvarphi_{q}^{\perp}}\tr K \vvarphi_{q}^{\perp}}}, \qquad 
&& \vvarphi_{p} = \frac{\vvarphi_{p}^{\perp}}{\sqrt{{\vvarphi_{p}^{\perp}}\tr M^{-1} \vvarphi_{p}^{\perp}}}.
\end{aligned}
\]
It it noteworthy that, in practice, the inverse of $M$ is never evaluated. Instead, one performs a Cholesky factorization to obtain the decomposition $M = L L\tr$. In particular, the normalization of $\vvarphi_{p}$ is done by forward and backward substitution, whose cost is negligible with respect to the overall complexity of the algorithm. Indeed, the main bottleneck is the factorization of $A_{q}$~\eqref{eq:matrixAq}, which needs to be performed at each iteration of the fixed point algorithm. We propose below two approaches that aim at:
\begin{itemize}
\item Reducing the number of iterations in the fixed point algorithm in order to reach convergence (see section~\ref{subsubsect:aitken});
\item Avoiding repetitive factorization of $A_{q}$ by carrying out computations in a subspace provided by the Ritz vectors, which are approximations of the eigenvectors of the generalized eigenproblem $K \uu = \lambda M \uu$ (see Section~\ref{subsect:ritz}).
\end{itemize}

\subsubsection{Problem in time}
\label{subsubsect:time_pgd}

We assume here that $\vvarphi$ is known and search for a new temporal mode $\ppsi$. We substitute $\zz_{m - 1} + \Phi \ppsi$ for $\zz$ in~\eqref{eq:condham_weak} and choose test functions in the form $\zzs = \Phi \ppsis$, with $\ppsis \in \mathcal Y^2$, where:
\[
\mathcal Y = C^0(\bar{\mathcal I}).
\]
Equation~\eqref{eq:condham_weak} reduces in this case to:
\[
\begin{aligned}
&& \int_{\Imega}{ {\ppsis}\tr \Phi\tr \left( J_{2 n} \Phi \ppsip + H \Phi \ppsi \right)~ dt } &= \int_{\Imega}{ {\ppsis}\tr \Phi\tr \left( \ff_{z} - J_{2 n} \zzp_{m - 1} - H \zz_{m - 1} \right)~ dt }, \qquad \forall \ppsis \in \mathcal Y^2,
\end{aligned}
\]
which simplifies to:
\[
\Phi\tr J_{2 n} \Phi \ppsip + \Phi\tr H \Phi \ppsi 
= \Phi\tr \left( \ff_{z} - J_{2 n} \zzp_{m - 1} - H \zz_{m - 1} \right).
\]
Above equation is discretized using the Crank-Nicolson time-marching scheme, such that, given $\ppsi^{0}$, one computes the $i$\textsuperscript{th} iterate ($i > 0$) as:
\begin{equation}
\label{eq:time_pgd}
A_{\mathcal{T}} \ppsi^{i} = B_{\mathcal{T}} \ppsi^{i - 1} + \bb_{\mathcal{T}}^{i}, 
\qquad i=1,\ldots,\nt
\end{equation}
where:
\[
\begin{aligned}
&A_{\mathcal{T}} = \begin{bmatrix}
h_{t} k_{x} & 2 c_{x} \\
- 2 c_{x} & h_{t} m_{x}
\end{bmatrix}, \\[.2in]
&B_{\mathcal{T}} = \begin{bmatrix}
- h_{t} k_{x} & 2 c_{x} \\
- 2 c_{x} & - h_{t} m_{x}
\end{bmatrix}, \\[.2in]
&\bb_{\mathcal{T}}^{i}
= \Phi\tr \Big[ h_{t} \Big( \ff_{z}^{i} + \ff_{z}^{i - 1} - H \left( \zz_{m - 1}^{i} + \zz_{m - 1}^{i - 1}\right) \Big) 
- 2 J_{2 n} \Big( \zz_{m - 1}^{i} - \zz_{m - 1}^{i - 1} \Big) \Big],
\end{aligned}
\]
and:
\[
\begin{aligned}
&k_{x} = \vvarphi_{q}\tr K \vvarphi_{q}, \\
&c_{x} = \vvarphi_{q}\tr \vvarphi_{p}, \\
&m_{x} = \vvarphi_{p}\tr M^{-1} \vvarphi_{p}.
\end{aligned}
\]
For each time step, Eq.~\eqref{eq:time_pgd} represents a $2 \times 2$ linear system that can be explicitly solved for $\ppsi^{n}$. Overall, the time problem is relatively cheap to solve as the cost is linear in the number of time steps~$n_{t}$. As previously mentioned, $\vvarphi_{q}$ and $\vvarphi_{p}$ are normalized after~\eqref{eq:space_pgd} is solved, so that $k_{x} = m_{x} = 1$ and only $c_{x}$ needs to be updated.

\subsection{Aitken acceleration} 
\label{subsubsect:aitken}

In the context of PGD order-reduced modeling, the number of iterations performed by the fixed-point algorithm has a direct impact on the efficiency of the approach. We propose here to employ the Aitken's $\Delta^2$ process to reduce the number of iterations that are necessary to reach convergence.

Let $\lin(\nxx)$ denote the complexity associated with solving one linear system of $\nxx$ algebraic equations in $\nxx$ unknown variables ($\lin(\nxx) \approx \mathcal O(\nxx^3)$ for fully-populated matrices). In structural dynamics simulations, the usual approach is to discretize the continuous equations with respect to the spatial variables using the finite element method and then obtain a system of $\nxx$ ordinary differential equations in the time variable $t \in \Imega$. The system is thereafter discretized in time by means of an (implicit) integration scheme (e.g.~Euler, Newmark, Crank-Nicolson, Hilber-Hughes-Taylor, \ldots). The degrees of freedom are then evaluated at each time step by solving a linear system of size $\nxx$. In the case of $n_{t}$ time steps, the complexity of the approach amounts to $n_{t} \lin(\nxx)$.

In the PGD framework, the solution of the problems in space and time is decoupled such that at each fixed-point iteration, one system of size $\nxx$ is solved for the spatial mode~\eqref{eq:space_pgd1} and one system of size two is solved $\nt$ times (marching scheme) for the temporal mode~\eqref{eq:time_pgd}. The complexity of one fixed-point iteration can be assumed to be of the order of $\lin(\nxx) + n_{t}$. It follows that the overall complexity of the PGD algorithm will be $m k_{\max} ( \lin( \nxx ) + n_{t} )$, where $m$ denotes the rank of the decomposition, i.e.\ the number of modes, and $k_{\max}$ is the maximal number of iterations allowed in the fixed-algorithm, whether or not convergence is reached. It can be inferred that a space-time separated PGD algorithm is competitive against a classical full-order solver whenever the following inequality holds:
\[
m k_{\max} ( \lin( \nxx ) + n_{t} ) \ll n_{t} \lin( \nxx ),
\]
highlighting the fact that the efficiency of the PGD algorithm highly depends on the number of fixed-point iterations.

The computation of an enrichment mode involves the following operators, formally written, at any given fixed-point iteration~$k$:
\begin{itemize}
\item $\mathcal{S}^{(k)} : \ppsi^{(k - 1)} \mapsto \vvarphi^{(k)}$, the operator that solves the system~\eqref{eq:space_pgd} for $\vvarphi^{(k)}$ with $\ppsi^{(k - 1)}$ given;
\item $\mathcal{T}^{(k)} : \vvarphi^{(k)} \mapsto \ppsi^{(k)}$, the operator that solves the system~\eqref{eq:time_pgd} for $\ppsi^{(k)}$ with $\vvarphi^{(k)}$ given.
\end{itemize}
As a result, the fixed-point algorithm computes two sequences $\big( \vvarphi^{(k)} \big)_{1 \leqslant k \leqslant k_{\max}}$ and $\big( \ppsi^{(k)} \big)_{1 \leqslant k \leqslant k_{\max}}$ until convergence. These sequences can be defined by recurrence relations as follows:
\[
\begin{aligned}
\vvarphi^{(k)} &= \mathcal{S}^{(k)} \circ \mathcal{T}^{(k - 1)} \big( \vvarphi^{(k - 1)} \big), \\
   \ppsi^{(k)} &= \mathcal{T}^{(k)} \circ \mathcal{S}^{(k)} \big( \ppsi^{(k - 1)} \big).
\end{aligned}
\]
The fixed-point convergence hinges upon the contraction property of the operators  $\mathcal{S}^{(k)} \circ \mathcal{T}^{(k - 1)}$ and $\mathcal{T}^{(k)} \circ \mathcal{S}^{(k)}$ for $\vvarphi^{(k)}$ and $\ppsi^{(k)}$ respectively. One common way to improve fixed-point iterations is by using relaxation techniques. This helps achieve a contraction property and usually enhances the convergence rate. The introduction of relaxation parameters $\omega_{\varphi}$ and $\omega_{\psi}$ leads to the following formulation of a fixed-point iteration:
\[
\begin{aligned}
\vvarphi^{(k)} &= \omega_{\varphi} ~\mathcal{S}^{(k)} \circ \mathcal{T}^{(k - 1)} \big( \vvarphi^{(k - 1)} \big) & \hspace{-.1in} + \left( 1 - \omega_{\varphi} \right) \vvarphi^{(k - 1)}, \\
   \ppsi^{(k)} &= \omega_{\psi} ~\mathcal{T}^{(k)} \circ \mathcal{S}^{(k)} \big( \ppsi^{(k - 1)} \big) & \hspace{-.1in} + \left( 1 - \omega_{\psi} \right) \ppsi^{(k - 1)}.
\end{aligned}
\]
In practice, it is preferable to adapt $\omega_{\varphi}$ and $\omega_{\psi}$ at each iteration. The so-called Aitken's delta square method provides a useful heuristic for determining the sequences $\omega_{\varphi}^{(k)}$ and $\omega_{\psi}^{(k)}$. One can also choose to enforce relaxation on the generalized coordinates modes and the conjugate momentum modes separately. In the Algorithm~\ref{algo:fixed_point_aitken}, Aitken acceleration is applied on the spatial modes only and separately for $\vvarphi_{q}$ and $\vvarphi_{p}$. Note that steps 15 and 16 of algorithm~\ref{algo:fixed_point_aitken} are not implemented in practice. Instead, space-time separation should be leveraged to efficiently compute stagnation coefficients in step 17.

\begin{algorithm}[tb]
\setstretch{1.6}
\caption{Fixed point algorithm with Aitken acceleration}
\label{algo:fixed_point_aitken}
\begin{algorithmic}[1]
\State \textbf{Initialization:} Set $k \leftarrow 0$
\State \phantom{\textbf{Initialization:}} Set $s_{q} \leftarrow \epsilon + 1$ and $s_{p} \leftarrow \epsilon + 1$ (with $\epsilon = 10^{-9}$)
\State \phantom{\textbf{Initialization:}} Set $\ppsi^{(0)}$ and $\vvarphi^{(0)}$
\While{$k < k_{\mathrm{max}}$ and $\max (s_{q}, s_{p}) > \epsilon$}
\State Increment the iteration counter: $k \leftarrow k + 1$
\State Compute new spatial modes: $\vvarphi \leftarrow \mathcal{S}^{(k)}\big( \ppsi^{(k - 1)} \big)$
\State Project modes: $\vvarphi_{q} \leftarrow P_{q} \vvarphi_{q}$ and $\vvarphi_{p} \leftarrow P_{p} \vvarphi_{p}$
\State Normalize modes: $\vvarphi_{q}^{(k)} \leftarrow \frac{\vvarphi_{q}}{\vvarphi_{q}\tr K \vvarphi_{q}}$ and $\vvarphi_{p}^{(k)} \leftarrow \frac{\vvarphi_{p}}{\vvarphi_{p}\tr M^{- 1} \vvarphi_{p}}$
\State Update spatial residual: $\rr_{q}^{(k)} = \vvarphi_{q}^{(k)} - \vvarphi_{q}^{(k - 1)}$ and $\rr_{p}^{(k)} = \vvarphi_{p}^{(k)} - \vvarphi_{p}^{(k - 1)}$
\If{$k > 1$}
\State Aitken $\Delta^{2}$: $\vvarphi_{q}^{(k)} \leftarrow \omega_{q}^{(k)} \vvarphi_{q}^{(k)} + \left( 1 - \omega_{q}^{(k)} \right) \vvarphi_{q}^{(k - 1)}$ with $\omega_{q}^{(k)} = \omega_{q}^{(k - 1)} \frac{{\rr_{q}^{(k - 1)}}\tr \left( \rr_{q}^{(k)} - \rr_{q}^{(k - 1)} \right)}{\left\| \rr_{q}^{(k)} - \rr_{q}^{(k - 1)} \right\|^{2}}$
\State \phantom{Relaxation:} $\vvarphi_{p}^{(k)} \leftarrow \omega_{p}^{(k)} \vvarphi_{p}^{(k)} + \left( 1 - \omega_{p}^{(k)} \right) \vvarphi_{p}^{(k - 1)}$ with $\omega_{p}^{(k)} = \omega_{p}^{(k - 1)} \frac{{\rr_{p}^{(k - 1)}}\tr \left( \rr_{p}^{(k)} - \rr_{p}^{(k - 1)} \right)}{\left\| \rr_{p}^{(k)} - \rr_{p}^{(k - 1)} \right\|^{2}}$
\EndIf
\State Compute new temporal mode: $\ppsi^{(k)} \leftarrow \mathcal{T}^{(k)}\big( \vvarphi^{(k)} \big)$
\State Compute: $\Delta_{q} \leftarrow \vvarphi_{q}^{(k)} \ppsi_{q}^{(k)} - \vvarphi_{q}^{(k - 1)} \ppsi_{q}^{(k - 1)}$ and $\Sigma_{q} \leftarrow \frac{1}{2} \left( \vvarphi_{q}^{(k)} \ppsi_{q}^{(k)} + \vvarphi_{q}^{(k - 1)} \ppsi_{q}^{(k - 1)} \right)$
\State \phantom{Compute:} $\Delta_{p} \leftarrow \vvarphi_{p}^{(k)} \ppsi_{p}^{(k)} - \vvarphi_{p}^{(k - 1)} \ppsi_{p}^{(k - 1)}$ and $\Sigma_{p} \leftarrow \frac{1}{2} \left( \vvarphi_{p}^{(k)} \ppsi_{p}^{(k)} + \vvarphi_{p}^{(k - 1)} \ppsi_{p}^{(k - 1)} \right)$
\State Evaluate the stagnation coefficients: $s_{q} \leftarrow \| \Delta_{q} \|_{L^{2}} / \| \Sigma_{q} \|_{L^{2}}$ and $s_{p} \leftarrow \| \Delta_{p} \|_{L^{2}} / \| \Sigma_{p} \|_{L^{2}}$
\EndWhile
\State \textbf{Return} the modes $\ppsi \leftarrow \ppsi^{(k)}$ and $\vvarphi \leftarrow \vvarphi^{(k)}$ 
\end{algorithmic}
\end{algorithm}

\subsection{Temporal update and symplectic structure}
\label{subsect:tempuptsympl}

Greedy algorithms generally incorporate an update procedure that consists in updating all the temporal modes for a given set of spatial modes. For a decomposition of rank $m$, the spatial modes can be conveniently stored in the matrix $S$ of size $2 n \times 2 m$, defined as:
\[
S = \begin{bmatrix}
\vvarphi_{1}^{q} & \ldots & \vvarphi_{m}^{q} & & 0 & \\
& 0 & & \vvarphi_{1}^{p} & \ldots & \vvarphi_{m}^{p}
\end{bmatrix} = \begin{bmatrix}
S_{q} & 0 \\ 0 & S_{p}
\end{bmatrix},
\]
while the temporal modes can be vertically stored in the time-dependent vector $\ppsi$ of size $2 m \times 1$, defined as:
\[
\ppsi = \begin{bmatrix}
\psi_{1}^{q} \\ \vdots \\ \psi_{m}^{q} \\ \psi_{1}^{p} \\ \vdots \\ \psi_{m}^{p}
\end{bmatrix},
\]
such that the decomposition of rank $m$ of $\zz$ reads:
\[
\zz_{m}(t) = S \ppsi(t) .
\]
The temporal update is performed by substituting $S \ppsi$ for $\zz$ in~\eqref{eq:condham_weak} and choosing test functions in the form $\zzs = S \ppsis$. Equation~\eqref{eq:condham_weak} thus reduces to:
\[
\int_{\Imega}{ {\ppsis}\tr S\tr \left( J_{2 n} S \ppsip + H S \ppsi \right)~ dt } 
= \int_{\Imega}{ {\ppsis}\tr S\tr \ff_{z} ~ dt }, 
\qquad \forall \ppsis \in \mathcal{Y}^{2 m},
\]
which can be rewritten in matrix form, with $\ff_{\psi} = S\tr \ff_{z}$, as:
\begin{equation}
\label{eq:matrix_update}
S\tr J_{2 n} S \ppsip + S\tr H S \ppsi = \ff_{\psi}. 
\end{equation}
Time discretization of the above equation using the Crank-Nicolson marching scheme leads to:
\begin{equation}
\label{eq:upt_pgd}
A_{\mathcal{U}} \ppsi^{i} = B_{\mathcal{U}} \ppsi^{i - 1} + \bb_{\mathcal{U}}^{i}, 
\qquad i=1,\ldots, \nt,
\end{equation}
with:
\[
\begin{aligned}
&A_{\mathcal{U}} = \begin{bmatrix}
h_{t} K_{x} & 2 C_{x} \\
- 2 C_{x}\tr & h_{t} M_{x}
\end{bmatrix}, \qquad
B_{\mathcal{U}} = \begin{bmatrix}
- h_{t} K_{x} & 2 C_{x} \\
- 2 C_{x}\tr & - h_{t} M_{x}
\end{bmatrix}, \\[.2in]
&\bb_{\mathcal{U}}^{i} = h_{t} S\tr \Big( \ff_{z}^{i} + \ff_{z}^{i - 1} \Big),
\end{aligned}
\]
and:
\[
\begin{aligned}
&K_{x} = S_{q}\tr K S_{q}, \\
&C_{x} = S_{q}\tr S_{p}, \\
&M_{x} = S_{p}\tr M^{-1} S_{p}.
\end{aligned}
\]
The orthonormalization of $( \vvarphi_{i}^{q} )_{1 \leqslant i \leqslant m}$ and $( \vvarphi_{i}^{p} )_{1 \leqslant i \leqslant m}$ with $K$ and $M^{-1}$, respectively, results in\break $K_{x} = M_{x} = I_{m}$.

The update procedure can be interpreted as projecting Hamilton's equations onto the subspace generated by the vectors of $S$. The system to be solved is governed by the Hamiltonian $g$ whose Hessian is $G = S\tr H S$. This Hessian can be interpreted as the rank-$2m$ reduced counterpart of the Hessian operator $H$, such that:
\[
g(\ppsi) = \frac{1}{2} \ppsi\tr G \ppsi - \ppsi\tr \ff_{\psi},
\]
and the full-order vector is given by $\zz \simeq \zz_{m} = S \ppsi$. Assuming that Hamiltonian~$g$ is canonical, the Hamilton's canonical equations of such a reduced-order system read:
\[
\ppsip = \nabla^{\gamma} g,
\]
where the symplectic gradient is given by:
\[
\nabla^{\gamma} g = J_{2 m} \nabla g = J_{2 m} \left( G \ppsi + \ff_{\psi} \right).
\]
It follows that the Hamilton's equations can be written as:
\[
\ppsip = J_{2 m} \left( G \ppsi - \ff_{\psi} \right).
\]
Multiplying both sides of this equation by $J_{2 m}$ (recall that $J_{2 m} J_{2 m} = - I_{2 m}$) and rearranging the terms leads to:
\[
J_{2 m} \ppsip + S\tr H S \ppsi = \ff_{\psi}.
\]
Recalling here Eq.~\eqref{eq:matrix_update}:
\[
S\tr J_{2 n} S \ppsip + S\tr H S \ppsi = \ff_{\psi},
\]
one observes that that the two equations are identical if and only if $S\tr J_{2\nxx} S = J_{2 m}$, i.e.~if $S$ is a symplectic mapping, according to the definition of the symplectic Stiefel manifold~\eqref{eq:def_stiefel}. However, in general, $S$ is not symplectic nor $g$ is a canonical Hamiltonian. The product $S\tr J_{2 n} S$ writes:
\[
S\tr J_{2 n} S = \begin{bmatrix}
0 & S_{q}\tr S_{p} \\
- S_{p}\tr S_{q} & 0
\end{bmatrix}.
\]
This suggests that the symplectic property could be enforced by biorthogonalization of $( \vvarphi_{i}^{q} )_{1 \leqslant i \leqslant m}$ and $( \vvarphi_{i}^{p} )_{1 \leqslant i \leqslant m}$, such that:
\[
S\tr J_{2 n} S = \begin{bmatrix}
0 & I_{m} \\ - I_{m} & 0
\end{bmatrix} = J_{2 m}.
\]
However, this property is not ensured in the current algorithm since we chose to orthogonalize $( \vvarphi_{i}^{q} )_{1 \leqslant i \leqslant m}$ and $( \vvarphi_{i}^{p} )_{1 \leqslant i \leqslant m}$ with respect to $K$ and $M^{-1}$, respectively. Yet, it can be enforced via a post-processing procedure. Let $P$ and $Q$ be two matrices of size $m \times m$ such that:
\[ 
\hat{S}_{q}\tr \hat{S}_{p} = I_{m}, \qquad \text{with}\
\hat{S}_{q} = S_{q} Q,\ \text{and}\
\hat{S}_{p} = S_{p} P.
\]
It follows that:
\begin{equation}
\label{eq:biorthogonalization}
(S_{q}Q)\tr S_{p} P = Q\tr S_{q}\tr S_{p} P = I_{m} 
\end{equation}
In other words, the matrices $Q$ and $P$ recombine the columns of $S_{q}$ and $S_{p}$ such that $\left( \hat{\vvarphi}_{i}^{q} \right)_{1 \leqslant i \leqslant m}$ and $\left( \hat{\vvarphi}_{i}^{p} \right)_{1 \leqslant i \leqslant m}$ form a biorthogonal system. We can readily conceive two approaches, among others, to enforce~\eqref{eq:biorthogonalization}:
\begin{itemize}
\item 
The LU factorization $S_{q}\tr S_{p} = LU$ allows one to define $Q = L\mtr$ and $P = U^{-1}$;
\item 
The Singular Value Decomposition $S_{q}\tr S_{p} = U \Sigma V\tr$ allows one to define $Q = U\mtr \Sigma^{- 1/2}$ and $P = V\mtr \Sigma^{- 1/2}$ ($\Sigma^{- 1/2}$ is defined as the diagonal matrix whose coefficients are given by the square root of the inverse of the singular values if different from zero, and zero otherwise).
\end{itemize}
We note that the two procedures are computationally efficient since they are performed on reduced matrices ($m \ll n$). Therefore, it is possible to construct a symplectic basis by post-processing the basis calculated by the PGD solver.

\subsection{Projection in Ritz subspace} 
\label{subsect:ritz}

As previously mentioned, the main bottleneck of the PGD solver is the solution of~\eqref{eq:space_pgd} that requires one to factorize the operator $A_{q}$ at each fixed-point iteration. Even though Aitken transformation does reduce the PGD solver time, the computational cost of the repeated factorization makes the solver prohibitively expensive when the dimension of the finite element space is large. 

We recall here the problem in space~\eqref{eq:space_pgd}, expressed now at a given fixed-point iteration indexed by parameter~$k$:
\[
A_{\mathcal{S}}^{(k)} \vvarphi^{(k)} = \bb_{\mathcal{S}}^{(k)}
\]
with:
\[
\begin{aligned}
A_{\mathcal{S}}^{(k)} &= \left[ \int_{\Imega}{ {\Psi^{(k - 1)}}\tr J_{2 n} \Psip^{(k - 1)} + {\Psi^{(k - 1)}}\tr H \Psi^{(k - 1)}~ dt } \right] = \begin{bmatrix}
k_{t}^{(k)} K & c_{t}^{(k)} I_{\nxx} \\
d_{t}^{(k)} I_{\nxx} & m_{t}^{(k)} M^{-1}
\end{bmatrix} \\[.2in]
\bb_{\mathcal{S}}^{(k)} &= \int_{\Imega}{ {\Psi^{(k - 1)}}\tr \left( \ff_{z} - J_{2 n} \zzp_{m - 1} - H \zz_{m - 1} \right)~ dt }
\end{aligned}
\]
where $m_{t}^{(k)}$, $k_{t}^{(k)}$, $c_{t}^{(k)}$ and $d_{t}^{(k)}$ are computed from the temporal modes $\psi_{q}^{(k - 1)}$ and $\psi_{p}^{(k - 1)}$, as defined in Section~\ref{subsubsect:space_pgd}. In particular:
\[
A_{q}^{(k)} = \left[ m_{t}^{(k)} k_{t}^{(k)} K - c_{t}^{(k)} d_{t}^{(k)} M \right]
\]
Therefore, at each fixed-point iteration, the weights associated with the stiffness and mass operators~$K$ and~$M$, respectively, have to be modified and a new factorization of $A_{q}^{(k)}$ needs to be obtained.

Although $A_{q}^{(k)}$ varies from one iteration to the next, its spectral content remains similar because the operator is derived from a linear combination of $K$ and $M$ (both remaining constant). The proposed approach takes advantage of the later observation and consists in projecting Eq.~\eqref{eq:space_pgd1} onto the subspace of approximated eigen-vectors, namely the Ritz vectors, which verify the following properties (with $m \leqslant r \ll n$):
\[
\left( \hat{\Lambda}, \hat{V} \right) \in \mathbb{R}^{r \times r} \times \mathbb{R}^{n \times r},\quad \text{such that} \quad 
\hat{V}\tr K \hat{V} = \hat{\Lambda},\ \text{and} \quad
\hat{V}\tr M \hat{V} = I_{r},
\]
where the Ritz values and the Ritz vectors are:
\[
\begin{aligned}
\hat{\Lambda} &= \mathrm{diag}\left( \hat{\lambda}_{1},\ \ldots\ , \hat{\lambda}_{r} \right), \\
\hat{V} &= \begin{bmatrix}
\hat{\vv}_{1}\ \ldots\ \hat{\vv}_{r}
\end{bmatrix}.
\end{aligned}
\]
We now introduce the mapping $R$:
\[
R = \begin{bmatrix}
\hat{V} & 0 \\ 0 & M \hat{V}
\end{bmatrix},
\]
and remark that $R \in S_{p}(2 r, 2 n)$, i.e.\ $R$ is a symplectic mapping. In other words, the structure of the equations presented above holds, which can be written in terms of $\hat{\zz} \in \mathbb{R}^{r}$, that satisfies $\zz = R \hat{\zz}$, and the Hamiltonian $\mathcal G$ defined as:
\[
\mathcal G(\hat{\zz}) = \frac{1}{2} \hat{\zz}\tr G \hat{\zz} - \hat{\zz}\tr \ff_{\hat{z}},
\]
with:
\[
G = R\tr H R = \begin{bmatrix}
\hat{\Lambda} & 0 \\ 0 & I_{r}
\end{bmatrix}, \qquad \ff_{\hat{z}} = R\tr \ff_{z}.
\]
For the Hamiltonian $\mathcal G$, the problem in space~\eqref{eq:space_pgd} using $\vvarphi^{(k)} = R \hat{\vvarphi}^{(k)}$ can thus be rewritten as:
\begin{equation}
\label{eq:ritz_space_pgd}
\hat{A}_{\mathcal{S}}^{(k)} \hat{\vvarphi}^{(k)} = \hat{\bb}_{\mathcal{S}}^{(k)}, 
\end{equation}
with:
\[
\begin{aligned}
\hat{A}_{\mathcal{S}}^{(k)} &= R\tr A_{\mathcal{S}}^{(k)} R = \left[ \int_{\Imega}{ {\Psi^{(k - 1)}}\tr J_{2 r} \Psip^{(k - 1)} + {\Psi^{(k - 1)}}\tr G \Psi^{(k - 1)} ~dt } \right] = \begin{bmatrix}
k_{t}^{(k)} \hat{\Lambda} & c_{t}^{(k)} I_{r} \\
d_{t}^{(k)} I_{r} & m_{t}^{(k)} I_{r}
\end{bmatrix}, \\[.2in]
\hat{\bb}_{\mathcal{S}}^{(k)} &= R\tr \bb_{\mathcal{S}}^{(k)},
\end{aligned}
\]
and~\eqref{eq:space_pgd1} becomes a diagonal system expressed as:
\begin{equation}
\left[ m_{t}^{(k)} k_{t}^{(k)} \hat{\Lambda} - c_{t}^{(k)} d_{t}^{(k)} I_{r} \right] \hat{\vvarphi}_{q}^{(k)} = \hat{\bb}_{q}^{(k)}. \label{eq:diag_space_pgd}
\end{equation}
The complexity of the spatial problem~\eqref{eq:space_pgd} is now linear in terms of the dimension~$r$ of the Ritz subspace. The number of Ritz vectors $r$ has to be chosen sufficiently high with respect to the expected rank $m$ of the PGD approximation. Depending on the external load, one can compute the Ritz vectors associated to the Ritz values corresponding to the frequency band of interest. Here, we chose to retain the Ritz vectors whose Ritz values have the lowest magnitudes, as conventionally performed in structural dynamics~\cite{rixen}.

\section{Numerical examples and discussion}
\label{sect:numerical_results}

\subsection{Test case: asymetric triangle wave Neumann boundary condition}

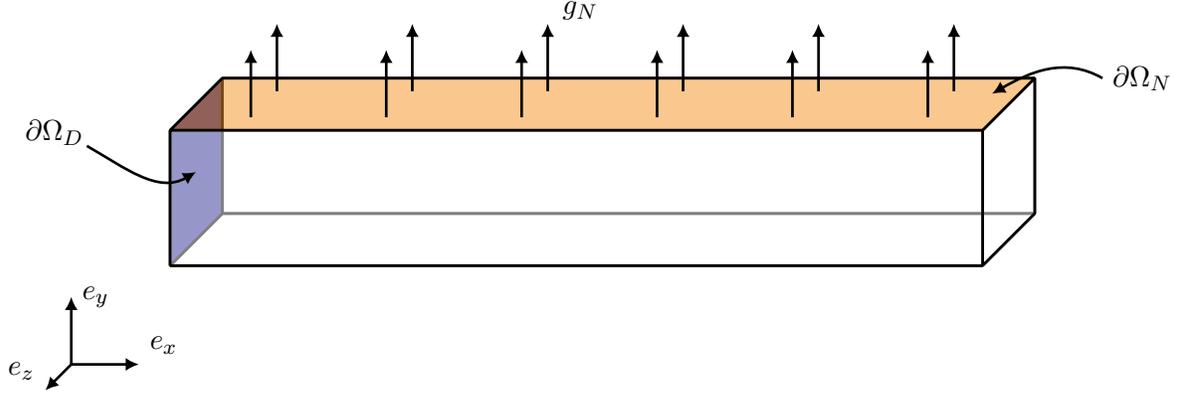
\begin{figure}[tb]
\centering
\begin{tikzpicture}[scale=1.8]
\coordinate (O) at (-1.5,-1.5,-1);
\coordinate (A) at (0,0,0);
\coordinate (B) at (6,0,0);
\coordinate (C) at (6,1,0);
\coordinate (D) at (0,1,0);
\coordinate (E) at (0,0,1);
\coordinate (F) at (6,0,1);
\coordinate (G) at (6,1,1);
\coordinate (H) at (0,1,1);
\coordinate (OD) at (-1., .5, 0);
\coordinate (ODl) at (-.95, .6, 0);
\coordinate (ON) at (6.5, 1., 0);

\draw[line width=1px, miter limit=1] (A) -- (B) -- (C) -- (D) -- cycle;
\draw[line width=1px, miter limit=1] (A) -- (E) -- (F) -- (B) -- cycle;
\draw[line width=1px, miter limit=1, fill=Blue] (A) -- (E) -- (H) -- (D) -- cycle;
\draw[line width=1px, miter limit=1, fill=white, fill opacity=.5] (B) -- (F) -- (G) -- (C) -- cycle;
\draw[line width=1px, miter limit=1, , fill=BurntOrange, fill opacity=.5] (D) -- (H) -- (G) -- (C) -- cycle;
\draw[line width=1px, miter limit=1, fill=white, fill opacity=.5] (E) -- (F) -- (G) -- (H) -- cycle;

\draw[-{Latex[length=5px, width=5px]}, line width=1px] (O) -- ++(.5,0,0) node[above right] {$e_{x}$};
\draw[-{Latex[length=5px, width=5px]}, line width=1px] (O) -- ++(0,.5,0) node[right] {$e_{y}$};
\draw[-{Latex[length=5px, width=5px]}, line width=1px] (O) -- ++(0,0,.5) node[above left] {$e_{z}$};

\draw[-{Latex[length=5px, width=5px]}, line width=1pt] (OD) to[out=-30, in=210] (0, .5, .5);
\draw (ODl) node[left] {$\partial\Omega_{D}$};

\draw[-{Latex[length=5px, width=5px]}, line width=1pt] (ON) to[out=-210, in=30] (5.8, 1, .3);
\draw (ON) node[right] {$\partial\Omega_{N}$};
\draw (2.85, 1.5, 0) node[left] {$g_{N}$};
\foreach \x in {0, 1, ..., 5} {
    \foreach \y in {.25, .75} {
        \draw[-{Latex[length=5px, width=5px]}, line width=1px] (\x+0.5, 1., \y) -- ++(0, 0.5, 0);
    }
}
\end{tikzpicture}
\caption{Scheme of the test case.}
\label{fig:3d_beam}
\end{figure}

\begin{figure}[tb]
    \centering
    \includegraphics[width=.6\textwidth]{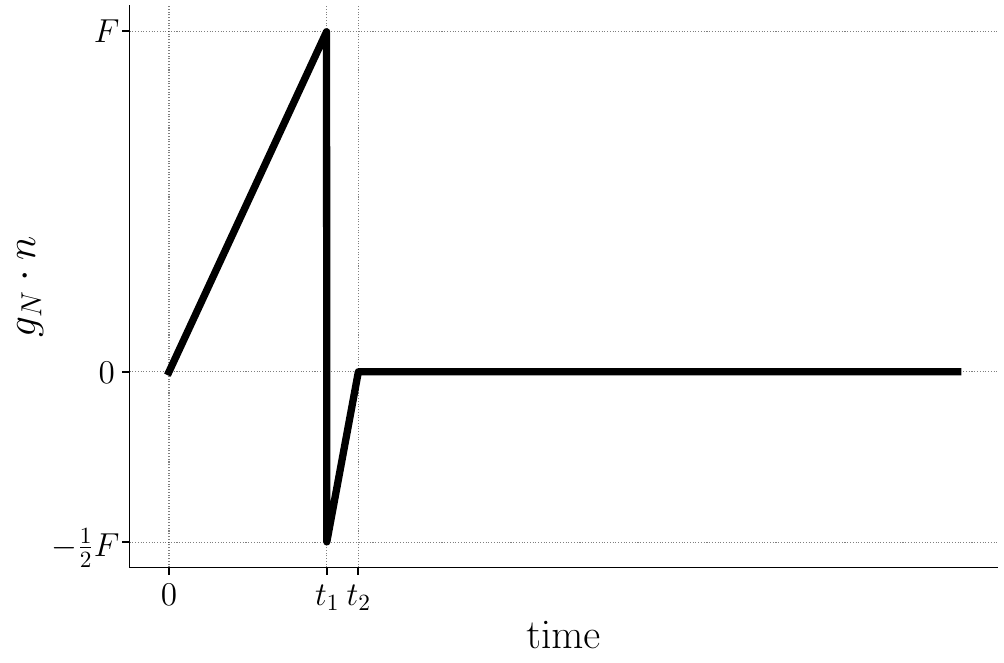}
    \caption{Evolution in time of the boundary traction $g_{N} \cdot n$ through time.}
    \label{fig:external_load}
\end{figure}

\begin{figure}[tb]
    \centering
    \includegraphics[width=.8\textwidth]{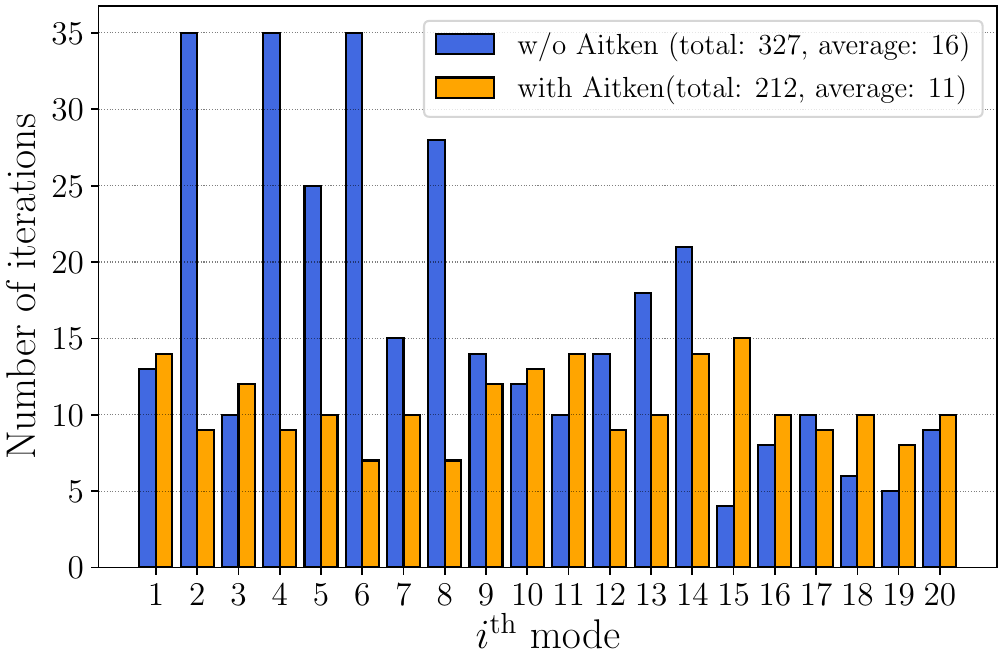}
    \caption{Number of iterations for 20 modes without and with Aitken acceleration}
    \label{fig:aitken}
\end{figure}

\begin{figure}[tb]
    \centering
    \includegraphics[width=0.75\textwidth]{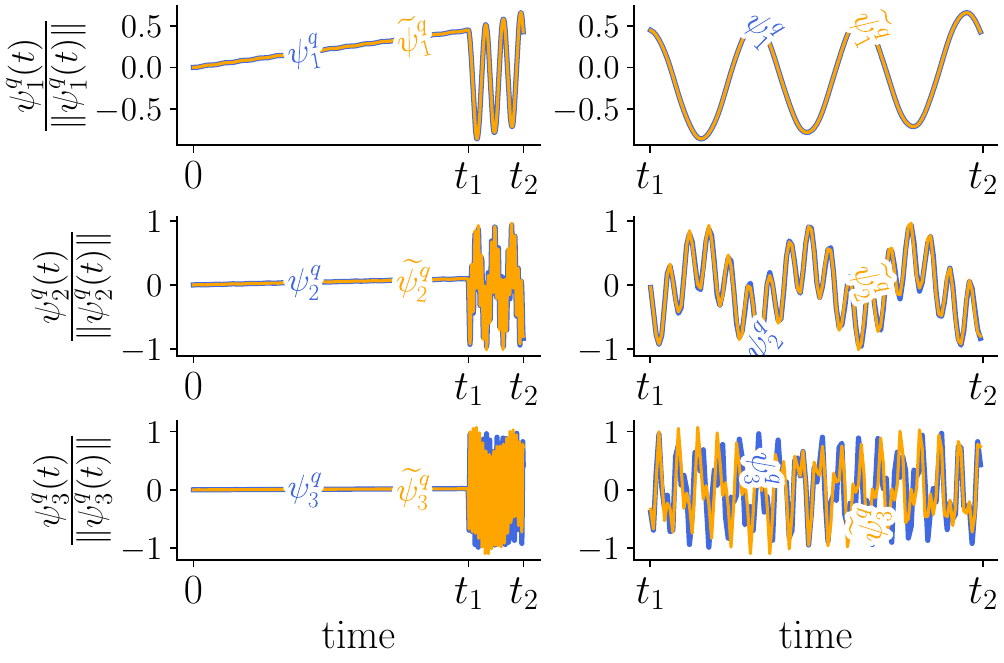}
    \caption{Visualization of the first three temporal modes (normalized) with and without Aiken acceleration, herein denoted $\widetilde{\psi}_{i}^{q}$ and $\psi_{i}^{q}$, respectively.}
    \label{fig:aitken_psi}
\end{figure}

The test case is inspired by an example found in~\cite{fischerPREPRINT} and has the interest of showcasing a transient phase followed by a steady-state harmonic regime. A 3D beam is considered, such that the domain $\Omega = (0, 6) \times (0, 1) \times (0, 1)$ (in meters) is a parallelepiped with a squared cross-section (see Figure~\ref{fig:3d_beam}). Its response to an external load on its top surface is computed in $\Imega = (0, 5)$ (in seconds):
\[
\rho \frac{\partial^{2} u}{\partial t^{2}} - \nabla \cdot \sigma(u) = 0, \qquad \forall (x, t) \in \Omega \times \Imega,
\]
with:
\[
\begin{aligned}
&\sigma(u) = 2 \mu \varepsilon( u ) + \lambda \mathrm{tr} \left( \varepsilon(u) \right) I_{3}, \\
&\varepsilon(u) = \frac{1}{2} \left( \nabla u + \left( \nabla u \right)\tr \right).
\end{aligned}
\]
Moreover, the beam is subjected to homogeneous initial conditions:
\[
\begin{aligned}
&u(x, 0) = 0, &&\quad \forall x \in \Omega, \\
&\frac{\partial u}{\partial t}(x, 0) = 0, &&\quad \forall x \in \Omega,
\end{aligned}
\]
and to the boundary conditions:
\[
\begin{aligned}
&u(x, t) = 0, &&\quad \forall (x, t) \in \partial \Omega_{D} \times \Imega, \\
&\sigma(u) \cdot n = g_{N}(x, t), &&\quad \forall (x, t) \in \partial \Omega_{N} \times \Imega,\\
&\sigma(u) \cdot n = 0, &&\quad \forall (x, t) \in \partial\Omega_{0} \times \Imega.
\end{aligned}
\]
In other words, the beam is clamped on its left end $\partial \Omega_{D} = \{0 \} \times (0, 1) \times (0, 1)$, an external load $g_{N} \cdot n$ is applied on its top surface $\partial \Omega_{N} = (0, 6) \times \{ 1 \} \times (0, 1)$ such that:
\[
g_{N}(x, t) = 
\left\{ 
\begin{aligned}
&\frac{t}{t_{1}} F, && \text{if}\ t < t_{1}, \\
- &\frac{1}{2} \left(1-\frac{t-t_{1}}{t_{2}-t_{1}}\right) F, && \text{if}\ t_{1} \leqslant t < t_{2}, \\
&0, && \text{otherwise},
\end{aligned} 
\right.
\]
where $t_{1} = 0.625$ and $t_{2} = 0.75$. In other words,
the external load pulls the beam upwards for $t \in [0, t_{1})$ and pushes it downwards for $t \in [t_{1}, t_{2})$ (see Figure~\ref{fig:external_load}). Finally, the beam is free on the remainder of the boundary $\partial\Omega_{0} = \partial\Omega \backslash (\partial \Omega_{D} \cup \partial \Omega_{N})$. \rev{In the space-discrete, time-continuous Hamiltonian formalism, the problem reads:
\[
\begin{aligned}
\ppp &= - K \qq + \ff, \\
\qqp &= M^{-1} \pp,
\end{aligned}
\]
with:
\[
\begin{aligned}
\qq(0) = 0, \\
\pp(0) = 0,
\end{aligned}
\]
where the stiffness and mass matrices, $K$ and $M$ respectively, result from the enforcement of the homogeneous Dirichlet boundary conditions by eliminating the corresponding rows and columns; the right-hand side $\ff$ is computed from the prescribed Neumann boundary conditions.}

The values of the parameters are chosen as follows:
\[
\begin{aligned}
&E = 220\ \text{GPa}, \\
&\nu = 0.3, \\
&\rho = 7000\ \text{kg/m\textsuperscript{3}}, \\
&F = 0.5\ \text{GPa},
\end{aligned}
\]
and the Lam\'e coefficients are evaluated as:
\[
\mu = \frac{E}{2 (1 + \nu)}, \qquad  \lambda = \frac{E \nu}{(1 + \nu)(1 - 2 \nu)}.
\]
The time domain $\Imega$ is divided into $\nt = 4800$ sub-intervals of equal size. The space domain $\Omega$ is partitioned into linear tetrahedral elements and five discretizations will be considered such that the number of spatial degrees of freedom (DOF) $2 \nx$ is chosen in $\{ 1~302, 6~204, 36~774, 67~032, 244~926 \}$. The number of Ritz vectors is set to $r = 300$ regardless of the spatial discretization. Unless otherwise stated, the reduced-order models are assessed on solutions involving $m = 50$ modes.

\subsection{Comparison method and performance criteria}

\begin{figure}[tb]
    \centering
    \includegraphics[width=0.80\textwidth]{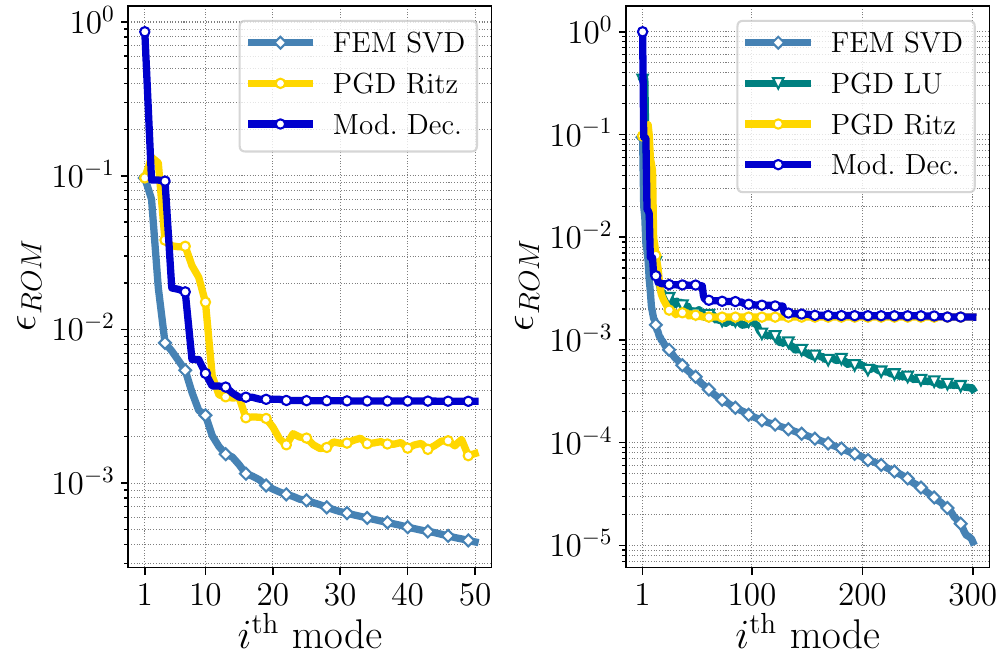}
    \caption{(Left) Error between the reference solutions and the SVD or PGD approximations for 244~926 spatial DOF with 50 modes. (Right) Error between the reference solutions and the SVD, PGD or Modal Decomposition approximations for 36~774 spatial DOF with 300 modes. ($y$-axis has log scale)}
    \label{fig:errors}
\end{figure}

\begin{figure}[tb]
    \centering
    \includegraphics[width=.8\textwidth]{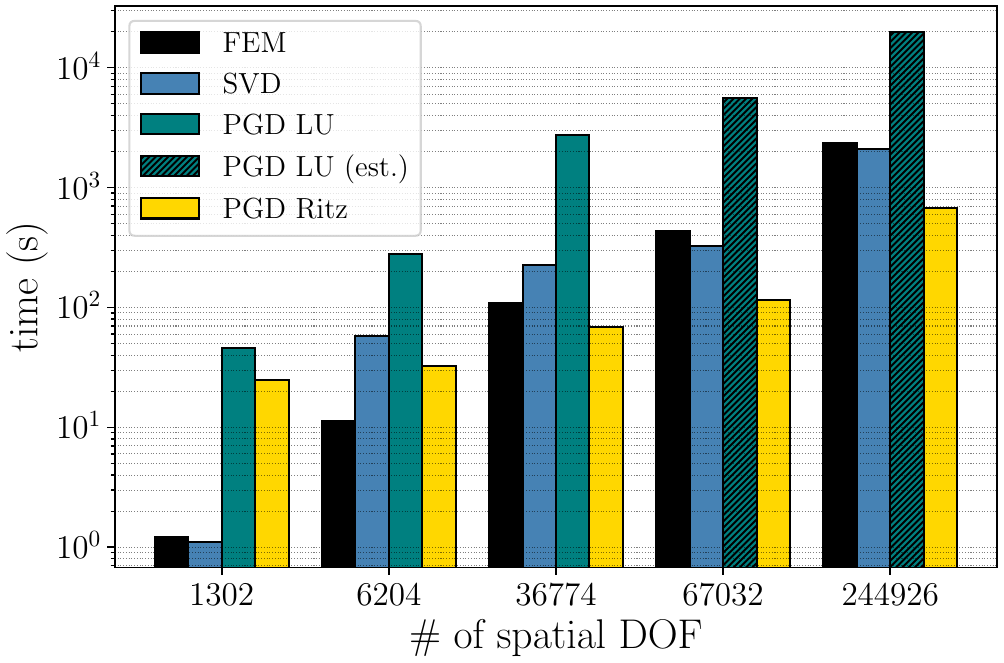}
    \caption{Real execution time for the full-order model (FEM) and the reduced-order models (SVD, PGD LU, PGD Ritz) with respect to different spatial discretizations ($y$-axis has log scale).}
    \label{fig:total_time}
\end{figure}

We shall report the results based on the following four features:
\begin{enumerate}[itemsep=0pt,topsep=4pt,parsep=0pt,leftmargin=25pt]
\item The number of fixed-point iterations without and with Aitken acceleration;
\item The accuracy of the PGD approximations with respect to full-order solutions, namely the FEM solutions described in Section~\ref{subsect:fem_ref};
\item The actual execution time of the different approaches and algorithms. The time efficiency of the PGD solvers will be detailed regarding the successive phases of the computation, namely the pre-processing, the fixed-point algorithm, the Gram-Schmidt algorithm, and the temporal update procedure.
\item The scalability of the approaches with respect to the size of the spatial discretization.
\end{enumerate}
The relative error $\epsilon_{\text{\itshape{ROM}}}$ in the reduced-order approximations with respect to the full-order solutions computed by the FEM is given by:
\[
\epsilon_{\text{\itshape{ROM}}} = \frac{\normi{u_\text{\itshape FEM} - u_\text{\itshape ROM} }}{\normi{u_\text{\itshape FEM}}}
\]
with \rev{$\normi{\cdot}$ being the energy norm:
\[
\normi{u} = \sqrt{ \int_{\Imega}{ \int_{\Omega}{ \frac{1}{2} \rho \dot{u} \cdot \dot{u} + \frac{1}{2} \epsilon(u) : \mathbb{E} : \epsilon(u) ~dx} dt} }.
\]
More precisely, in the space-discrete Hamiltonian framework, the energy norm will be evaluated as follows:
\[
\normi{u} = \sqrt{ \int_{\Imega}{ \frac{1}{2} \pp\tr M^{-1} \pp + \frac{1}{2} \qq\tr K \qq ~dt } }.
\]}%
Note that the full-order solution computed by the FEM is obtained using the same discretization parameters.

The reduced-order approximations that will be considered are the Singular Value Decomposition (SVD) of the full-order solution, the PGD LU that factorizes the space operator by LU decomposition for each fixed-point iteration and the PGD Ritz that computes the reduced-order model in the subspace spanned by the Ritz vectors. More precisely, we will present the errors with respect to the number of modes $m$ in the PGD solutions and compare these errors to those obtained by subsequently performing an SVD on the full-order solutions.

As far as computer times are concerned, all computations were run on a computer with the following configuration:
\begin{itemize}
\item CPU: AMD Ryzen 7 PRO 4750U @ 1.7 GHz per core (8 cores, 16 threads);
\item RAM: 38 GB;
\item OS: Arch Linux.
\end{itemize}
The code was written using Python 3.9.17 with NumPy 1.25.0~\cite{harris} and SciPy 1.10.1~\cite{scipy} built from sources and linked against BLAS/LAPACK and SuiteSparse~\cite{chen}.

\begin{figure}[tp]
    \centering
    \includegraphics[width=.8\textwidth]{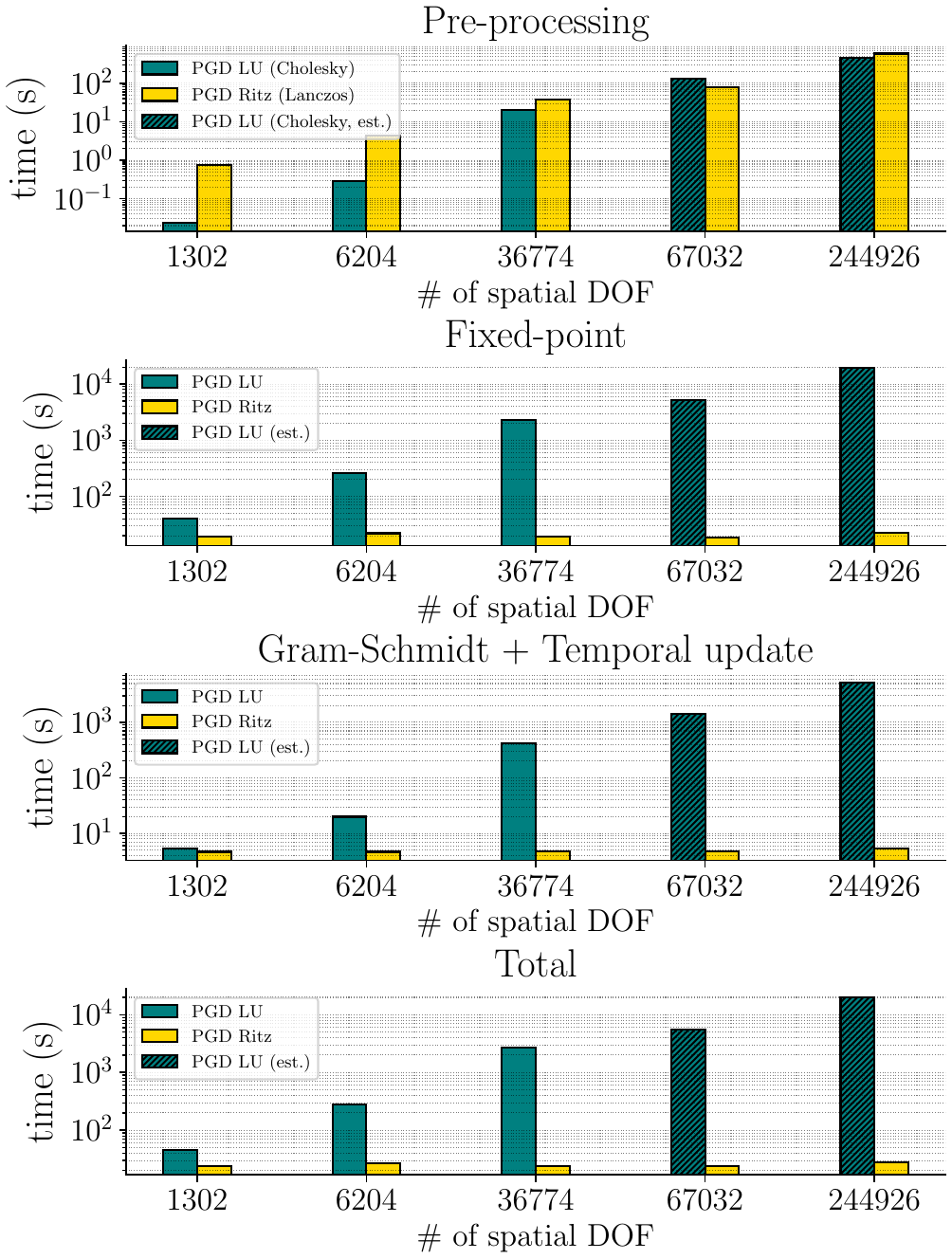}
    \caption{Detail on the real execution time for the full-order model (FEM) and the reduced-order models (SVD, PGD LU, PGD Ritz) with respect to different space discretizations ($y$-axis has log scale).}
    \label{fig:detail_time}
\end{figure}

\subsection{Numerical results}

\paragraph{Aitken acceleration.}

The relaxation technique significantly reduces the number of fixed-point iterations (see Figure~\ref{fig:aitken}). For 20 modes, Aitken acceleration saves five iterations per enrichment, on average, and a total of over 100 iterations for the full computation. Moreover, it is worth highlighting that without Aitken acceleration, the fixed-point procedure sometimes terminates without reaching convergence. This is the case for example for modes 2, 4, and 6, as shown in Figure~\ref{fig:aitken}. Indeed, the maximum number of iterations in this example is set to 35 iterations, so that if convergence is not reached within the 35 iterations, the fixed-point procedure is aborted and the last computed mode is retained. Thus, not only Aitken acceleration increases the computational efficiency, but also allows one to reach the convergence criterion that may not be satisfied otherwise. \rev{Eventually, slight discrepancies in the temporal modes may be noticeable between the results obtained with and without acceleration (see Figure~\ref{fig:aitken_psi}). On the other hand, there is no significant difference on the spatial modes, as illustrated in Figure~\ref{fig:spatial_modes}, with Aitken acceleration when using either one of the two PGD approaches.}

\paragraph{ROM accuracy.}

Figure~\ref{fig:errors} shows the errors of the reduced-order models with respect to the FEM solutions for $2n = $ 244,926 spatial degrees of freedom. We observe that the errors significantly decrease for both the PGD LU and PGD Ritz approaches during the 20 first modes. In fact, the accuracy of the PGD Ritz solution is similar to that of the PGD LU solutions. Moreover, we observe that the convergence of the two PGD approximations is comparable to that of the SVD, at least for the 20 first modes, before reaching a plateau.

\paragraph{Execution time and scalability.}

Figures~\ref{fig:total_time} and~\ref{fig:detail_time} show respectively the total and detailed real execution times of the different methods. We remark that the PGD solver is not competitive when the number of degrees of freedom remains low. We also observe that, except in the case with 1,302 spatial degrees of freedom, the PGD Ritz outperforms any other method. On the one hand, the SVD, as an \textit{a posteriori} method, requires a full-order snapshot to build a reduced-order model. Moreover, the extraction of the principal components from the data takes as much time as the actual full-order computation. On the other hand, the Ritz version of the PGD solver as an \textit{a priori} method does not require any prior knowledge of the full-order solution and reaches an error comparable to that of the SVD for the first 20 modes. More precisely, the PGD Ritz does not reach an error as low as that of the SVD. However, the difference in error is small enough in view of the speedup to justify the use of the PGD Ritz over the SVD (see Table~\ref{tab:femsvdvspgd}). Conversely, the use of the PGD LU in this context cannot really be justified over the SVD.

Regarding the detailed execution times, it seems that the pre-processing phase has comparable computational efficiency. In other words, the computation of a Cholesky factorization for $M$ is as costly as computing Ritz pairs. Nevertheless, carrying out the PGD computation in the subspace provided by the Ritz vectors drastically increases the performance of each of the subsequent phases, namely the fixed-point, Gram-Schmidt, and the temporal update procedures.

\begin{table}[tb]
\setstretch{1.6}
\centering
\caption{Time efficiency of the reduced-order models (SVD, PGD LU, PGD Ritz) with respect to different spatial discretizations and PGD Ritz speedup compared to other methods.}
\label{tab:femsvdvspgd}
\begin{tabular}{|c|c|c|c|c|c|}
\hline 
\textbf{\# DOF in space}
& $\begin{subarray}{c} \Delta T_{1} ~\textbf{(s)} \\ \textbf{(FEM \& SVD)} \end{subarray}$
& $\begin{subarray}{c} \Delta T_{2} ~\textbf{(s)} \\ \textbf{(PGD LU)} \end{subarray}$
& $\begin{subarray}{c} \Delta T_{3} ~\textbf{(s)} \\ \textbf{(PGD Ritz)} \end{subarray}$
& \textbf{Gain $\frac{\Delta T_{1}}{\Delta T_{3}}$}
& \textbf{Gain $\frac{\Delta T_{2}}{\Delta T_{3}}$} \\
\hline 
1,302 & 2.32 & 46.23 & 24.95 & 0.09 & 1.85 \\
\hline
6,204 & 68.99 & 280.69 & 32.38 & 2.13 & 8.67 \\
\hline
36,774 & 333.26 & 2,750.84 & 68.77 & 4.85 & 40.00 \\
\hline
67,032 & 760.15 & n.a. & 115.79 & 6.57 & n.a. \\
\hline
244,926 & 4,428.65 & n.a. & 676.87 & 6.54 & n.a. \\
\hline
\end{tabular}
\end{table}

\subsection{Further discussion}

\begin{figure}[tb]
    \centering
    \includegraphics[width=\textwidth]{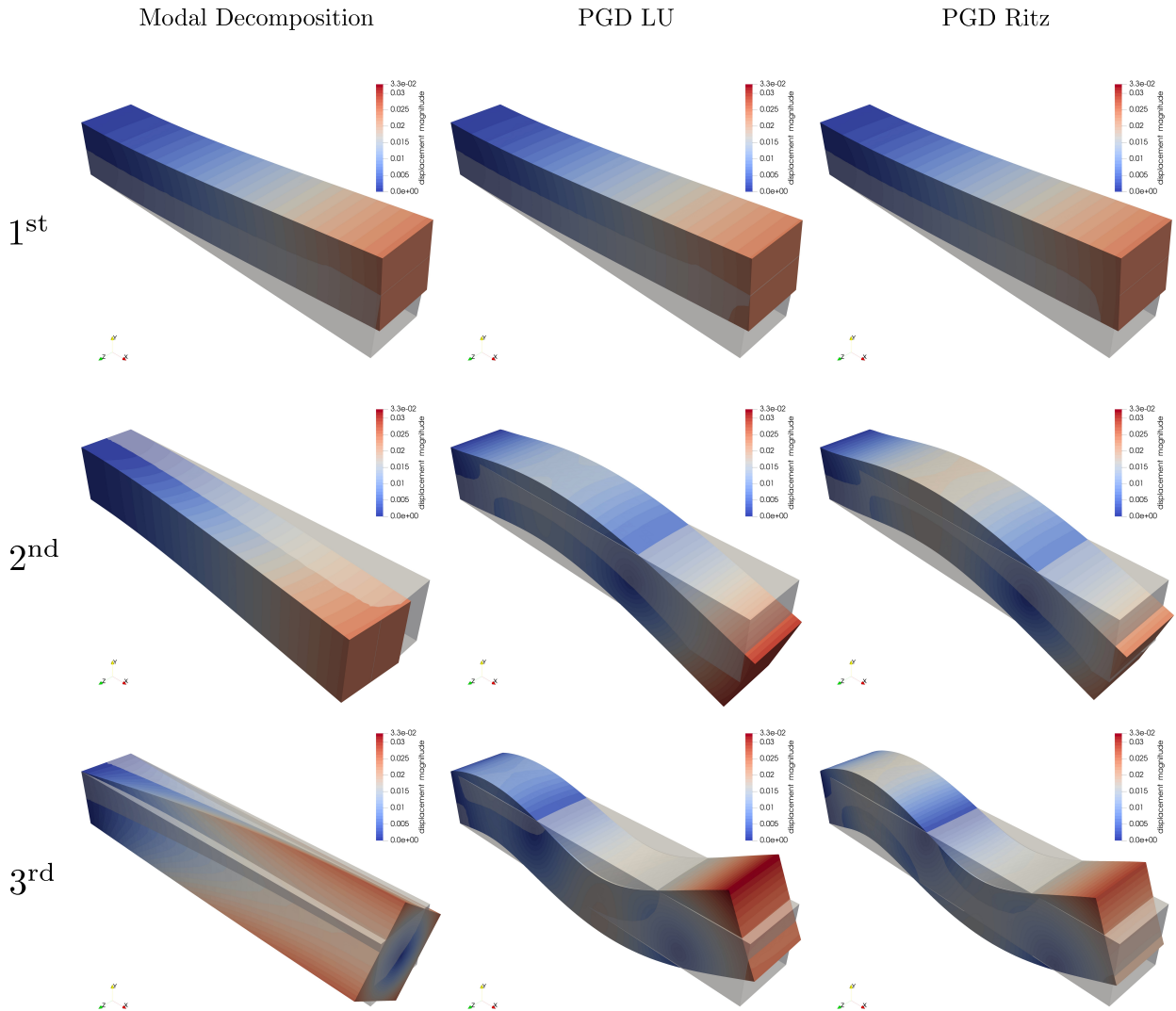}
    \caption{Visualization of the first three spatial modes (normalized) for the Modal Decomposition, PGD LU and PGD Ritz on the first, second and third columns, respectively and undeformed configuration in low opacity.}
    \label{fig:spatial_modes}
\end{figure}

The PGD Ritz solver is overall much more efficient than the other approaches and offers a remarkably good compromise in terms of error decay. Moreover, this novel approach displays good scalability with respect to the number of spatial degrees of freedom, with a reasonable error for a relatively small number of modes, which is highly suitable in model-order reduction. The PGD Ritz solver could be interpreted as a hybrid approach between classic PGD solvers and Modal Decomposition methods. In that respect, the relevance of the PGD Ritz over classic PGD solvers is unequivocal in a space-time separated context. Yet, its advantage over Modal Decomposition must be discussed, as well as its potential to perform well if separation with additional parameters (material, geometric, etc.) had to be accounted for.

Around the 20\textsuperscript{th} mode, we observe on Figure~\ref{fig:errors} that the error decay slows down or even stagnates for the PGD Ritz. Since computations are carried out in the subspace spanned by the Ritz vectors, it is intuitively understandable that the quality of the PGD approximation is bounded by the information contained in the Ritz vectors. Indeed, Figure~\ref{fig:errors} illustrates this idea: the error in the solution obtained by the PGD Ritz after the first 20 enrichments matches the error of the response computed by Modal Decomposition (MD) with $r = 300$ modes (number of Ritz vectors). On the one hand, Ritz vectors are describing the natural response of the system. Thus, not all the Ritz vectors will be relevant to describe the structural response under external loads. Mode participation factors or methods such as sensitivity analysis or mode shape analysis may provide insights to select a set of vectors that capture a given dynamic behavior. However, these approaches can be tedious as they may require user intervention to interpret the results, which makes the process subjective and less repeatable. On the other hand, the PGD solver inherently accounts for external loads to compute relevant modes that describe the structural response accurately. In the PGD Ritz framework, it translates to find linear combinations of the Ritz vectors that satisfy the PGD spatial formulation~\eqref{eq:space_pgd} that derives from the Galerkin finite element formulation. This is well illustrated by Figure~\ref{fig:spatial_modes}: the first three modes for Modal Decomposition are the dominant deformation modes for the beam geometry, respectively vertical bending, lateral bending, and torsion. However, for the given external load, lateral bending and torsion are not relevant. We can thus see that, like modal decomposition, the PGD solvers compute a first mode corresponding to vertical bending, but the subsequent modes also contribute to the description of vertical bending, which is effectively the dominant mode to describe the structure's response to the prescribed load.

Figure~\ref{fig:errors} also shows that while the error in the PGD Ritz solution reaches a plateau, that in the PGD LU solution eventually keeps decreasing when the number of modes is increased. Thus, if error stagnation is detected while the accuracy remains above a given tolerance, two strategies can be considered:
\begin{itemize}
\item Restarting the Arnoldi algorithm to find subsequent Ritz vectors (i.e.~increase~$r$), so as to enrich the research space for new PGD modes;
\item Switch back to the PGD LU algorithm.
\end{itemize}

The methodology can be straightforwardly extended to viscoelastic systems modeled with Rayleigh damping, allowing for the construction of a parametric reduced-order model with respect to the Rayleigh damping coefficients. Indeed, the damping term does not change the matrix pattern of the system~\eqref{eq:space_pgd} to be solved for the spatial mode. In~\cite{cavaliere2022}, the parametric eigenproblem $K_{\mu} \uu(\mmu) = \lambda(\mmu) M_{\mu} \uu(\mmu)$ is solved for applications in structural dynamics, where the stiffness $K_{\mu}$ and mass $M_{\mu}$ operators depend on material or geometric parameters $\mmu$. The authors introduce an original method to solve this parametric eigenproblem within the PGD framework, so as to find approximations of the eigen-pairs $\left( \lambda(\mmu), \uu(\mmu) \right)$ in a parameter-separated format. Their approach may be considered to provide a parametrized subspace, onto which the spatial problem~\eqref{eq:space_pgd} can be projected to recover a diagonal structure as in~\eqref{eq:diag_space_pgd}. The PGD Ritz would optimize the selection of the eigenvectors that contribute to the structure response. Therefore, the PGD Ritz could present a proficient tool to compute the dynamic response of structures subjected to time-dependent loads, even in a parametrized setting. 

\rev{Furthermore, the possibility to choose a symplectic time integrator in combination with the preservation of symplecticity of the spatial modes offers an appropriate foundation for a potential extension of this work. It may allow for the development of a reduction technique suited to the treatment of elastodynamics problems that involve large rotations and small strains as presented in~\cite{simo}.} Finally, the proposed approach allows one to consider the construction of a PGD Ritz aimed at minimizing an error with respect to a Quantity of Interest (QoI)~\cite{kergrene}. The PGD subproblems would be modified so that combinations of the Ritz vectors are now sought for as to minimize a residual over a QoI. These topics will be studied in future works.

\section{Conclusion}
\label{sect:conclusions}

The PGD solver developed here combines good accuracy and efficiency, even with an increased number of degrees of freedom. The calculation of the PGD modes in the subspace spanned by the Ritz vectors proves to be proficient, as it substantially accelerates the computation without introducing a significant approximation error. Aitken acceleration and the orthogonalization procedures are not as important for computational efficiency, but guarantee convergence and stability properties that are essential to the solver. In addition, the solver, which is based on the Hamiltonian formalism, builds reduced models for both the generalized coordinates and conjugate momenta. It has been shown that it allows the construction of a symplectic reduced basis, thus respecting the structure of canonical Hamiltonian mechanics. This is an interesting feature, as it opens up a variety of avenues related to this fundamental structure in dynamics. The numerical results also show great promise regarding the viability of this approach for solving linear elastodynamics problems on three-dimensional structures.

\section{Declarations}

\subsection{Availability of data and materials}

All data generated or analysed during this study are included in this published article or available from the corresponding author on reasonable request.

\subsection{Competing interests}

The authors declare that they have no known competing financial interests or personal relationships that could have appeared to influence the work reported in this paper.

\subsection{Funding}

Cl\'ement Vella acknowlegdes the funding awarded by the French Ministry of National Education, Higher Education, Research and Innovation referred to as ``Specific Doctoral Contracts for Normaliens''. Serge Prudhomme is grateful for the support by a Discovery Grant from the Natural Sciences and Engineering Research Council of Canada [grant number RGPIN-2019-7154].

\subsection{Authors' contributions}

\begin{tabularx}{\textwidth}{@{}lX}
\textbf{Cl\'ement Vella}: & Conceptualization, Methodology, Software, Validation, Formal analysis,
Investigation, Writing -- original draft, Visualization. \\
\textbf{Pierre Gosselet}: & Conceptualization, Methodology, Formal analysis, Writing -- review \& editing.\\
\textbf{Serge Prudhomme}: & Conceptualization, Formal analysis, Writing -- review \& editing, Supervision, Funding acquisition, Discussions.
\end{tabularx}


\subsection{Acknowledgements}

Not applicable.

\appendix
\rev{%
\section{Time operators} \label{appendix:timeop}

The computation of time integrals is required to evaluate the coefficients of the problem in space presented in Section~4.1.1, i.e.~$k_{t}$, $c_{t}$, $d_{t}$, and $m_{t}$. Let $u = u(t)$ and $v = v(t)$ be two functions of time and assume they are sufficiently regular. We consider continuous, piecewise linear approximations of $u$ and $v$, which read in the case of $u$, and in a similar manner for $v$:
\[
u(t) \simeq \left( 1 - \frac{t - t^{i - 1}}{h_{t}} \right) u\left( t^{i - 1} \right) + \frac{t - t^{i - 1}}{h_{t}} u\left( t^{i} \right),\quad t \in \left[ t^{i - 1}, t^{i} \right],\ i=1, \ldots, \nt,
\]
with $h_{t} = t^{i} - t^{i - 1}$. We can now define the vectors $\uu, \vv \in \mathbb{R}^{\nt}$ as:
\[
\begin{aligned}
& \uu = \begin{bmatrix} u(t^{0}) & \ldots & u(t^{\nt}) \end{bmatrix}\tr, \\
& \vv = \begin{bmatrix} v(t^{0}) & \ldots & v(t^{\nt}) \end{bmatrix}\tr. 
\end{aligned}
\]
The time integrals are then approximated as:
\[
\begin{aligned}
\int_{\Imega}{u v ~dt} &\simeq \uu\tr A_{t} \vv, \\
\int_{\Imega}{\dot{u} v ~dt} &\simeq \uu\tr C_{t} \vv,
\end{aligned}
\]
with $A_{t}$ and $C_{t}$ the time operators such that:
\[
\begin{aligned}
A_{t} &= \frac{h_{t}}{6} &&\left[ \begin{tabular}{ccccc}
2 & 1 &  &  &  \\
& 4 & $\ddots$ & \phantom{\text{sym.}}[0] & \\
& & $\ddots$ & $\ddots$ & \\
& \phantom{[0]}\text{sym.} & & 4 & 1 \\
& & & & 2 \\
\end{tabular} \right], 
\\[.2in]
C_{t} &= \frac{1}{2} &&\left[ \begin{tabular}{ccccc}
-1 & -1 &  &  &  \\
1 & 0 & $\ddots$ & \phantom{\text{sym.}}[0] & \\
& $\ddots$ & $\ddots$ & $\ddots$ & \\
& [0]\phantom{\text{sym.}} & $\ddots$ & 0 & -1 \\
& & & 1 & 1 \\
\end{tabular} \right].
\end{aligned}
\]
}

\newpage
\bibliographystyle{abbrv}
\bibliography{literature.bib}

\end{document}